\documentclass[a4,12pt]{article} 
\usepackage{graphics}
\usepackage{latexsym}
\usepackage{amsbsy}
\newcommand{\PA}{{\mathcal{P}}}
\newcommand{\FU}{{\mathcal{F}}}
\newcommand{\N}{{N_{\epsilon}}}
\newcommand{\mathR}{{\rm I\! R}}
\newcommand{\pa}{ I^-(q)}
\newcommand{\fu}{I^+(q)}
\newcommand{\cpa}{\downarrow I^+(q)}
\newcommand{\cfu}{\uparrow I^-(q)}
\newcommand {\be}{\begin{equation}} 
\newcommand {\ee}{\end{equation}}
\newcommand {\bea}{\begin{eqnarray}} 
\newcommand{\eea}{\end{eqnarray}}

\newcommand{\pd}{\partial}

\newcommand{\ze}{\rm Z}

\newtheorem{corollary}{Corollary}
\newtheorem{proposition}{Proposition} 
\newtheorem{lemma}{Lemma}
\newtheorem{claim}{Claim}

\title{Causal continuity in degenerate spacetimes}
\author{A.Borde${}^{a,1}$, H.F.Dowker${}^{b, 2}$, R.S.Garcia${}^{c,2}$, R.D.Sorkin${}^{d,3}$, S.Surya${}^{e,4}$\\
        $\;$ \\
        ${}^1$  Institute of Cosmology, Department of Physics and 
        Astronomy\\ 
        Tufts University, Medford, MA 02155, USA \\
        ${}^2$ Blackett Laboratory,
         Imperial College of Science Technology and Medicine,\\
        London SW7 2BZ, United Kingdom \\
	${}^3$ Syracuse University, Syracuse, N.Y., 13244-1130, USA \\
	${}^4$ TIFR, Homi Bhabha Rd, Mumbai 400 005, India.}
\begin{document}
\begin{titlepage}
\maketitle
\begin{abstract} 

\thispagestyle{empty}

A change of spatial topology in a causal, compact spacetime cannot 
occur when the metric is globally Lorentzian. One can however
construct a causal metric from a Riemannian metric and a Morse
function on the background cobordism manifold, which is Lorentzian 
almost everywhere except that it is degenerate at
each critical point of the function.  
We investigate
causal structure in the neighbourhood of such a degeneracy, when the
auxiliary Riemannian metric is taken to be Cartesian flat in
appropriate coordinates.  For these geometries, we verify Borde and
Sorkin's conjecture that causal discontinuity occurs if and only  
if the Morse index is $1$ or $n-1$.
\end{abstract}
\vspace{-0.25 cm}
\noindent ${}^a$ {\small Email: borde@cosmos2.phy.tufts.edu.
        Permanent address: Theoretical and 
        Computational Studies Group,
        Natural Sciences Division, 
        Southampton College of Long Island University,
        Southampton, NY 11968, USA}\\
\noindent ${}^b$ {\small dowker@ic.ac.uk}\\
\noindent  ${}^c$ {\small garciars@ic.ac.uk, Beit Fellow}\\ 
\noindent ${}^d$ {\small sorkin@suhep.syr.edu} \\
\noindent ${}^e$ {\small ssurya@tifr.res.in} \\ 
\end{titlepage}

\section{Introduction}

Spatial topology change is incompatible with both a non-degenerate
Lorentzian metric and a causal partial order on all spacetime points
\cite{geroch67}. One or other of these conditions must be given up if
topology change is to occur. However, even if the causal order is abandoned
and closed timelike curves (CTCs) allowed, there are certain topology
changes, including physically interesting ones such as the pair production
of Kaluza-Klein monopoles, that still cannot occur via a globally
Lorentzian spacetime \cite{sorkin86a,sorkin86b}.  On the other hand if CTCs
are excluded, all possible topology changes are permitted at a kinematical
level as long as the metric is allowed to degenerate to zero at finitely
many isolated ``Morse'' points \cite{yodzis73,sorkin90}.

In the Sum-Over-Histories (SOH) framework for quantum gravity, the
transition amplitude between two non-diffeomorphic spacelike hypersurfaces
$V_0$ and $V_1$ is given as a sum over all interpolating geometries (see
for example
\cite{hawking78,iihawking78,sorkin86a,horowitz91,gibbons92,vilenkin94,carlip97}
). These geometries are usually taken to be globally defined. As discussed
in \cite{surya97,dowgar}, this prescription can be generalised to include the
so-called ``Morse geometries'' on the interpolating cobordism $M$, where
$\partial M=V_0\amalg V_1$ ($\amalg$ denoting disjoint union).  A 
Morse metric, $g$, is defined to be 
\be
\label{mmetric.eq}
g_{\mu\nu} \equiv h_{\mu\nu}(h^{\lambda\sigma} \partial_\lambda f
\partial_\sigma f) - \zeta \partial_\mu f \partial_\nu f \ee 
where $\zeta$ is a real number greater than one, $h$ is a Riemannian
metric on $M$ and $f$ is a Morse function $f:M\rightarrow [0,1]$ such
that $f^{-1}(0) = V_0$ and $f^{-1}(1) = V_1$. A Morse function, $f
\in C^{\infty}(M)$ has critical points, at which  $\partial_{\mu}f=0$, which
are isolated and nondegenerate, i.e. the Hessian of f is non-singular.
This metric is Lorentzian throughout the regular set of the Morse function
and vanishes at its critical points $\{p_k\}$. The index, $\lambda_k$, of
$p_k$ is the number of negative eigenvalues of the Hessian of $f$ at $p_k$.
We define a {\it Morse geometry} as the pair $(M,g)$, where $M$ is a
compact cobordism and $g$ a Morse metric on it.  We suggest that these
Morse geometries should in fact be included in the SOH for quantum gravity.

Motivated by surgery theory, Borde and Sorkin conjectured that only those
spacetimes which contain critical points of index $1$ or $n-1$ have causal
discontinuities \cite{BordeSorkin,surya97}. This conjecture, while on one
hand of mathematical interest, in fact has potential importance in quantum
gravity. It was shown by the authors of \cite{dewitt,iidray} that scalar
quantum field propagation on the $1+1$ trousers is singular and hence it
was suggested that such a topology will be suppressed in the
SOH. Subsequently, the authors of \cite{louko97} found that causally
discontinuous topology changing processes in $1+1$ dimensions are indeed
suppressed, while causally continuous ones are enhanced.  This led Sorkin
to further conjecture that singular propagation of quantum fields on such
backgrounds is related to the causal discontinuity of the spacetime.
    
It is the purpose of this paper to investigate the first of these, the
Borde-Sorkin conjecture. The standard analyses of causal structure however,
posit the existence of a globally Lorentzian metric
\cite{penrose72,hawking73,sachs73}, and thus exclude Morse geometries with
their isolated degeneracies.  However, we may investigate the geometries
induced in the regular set of the Morse function.  Thus we define a {\it
Morse spacetime} as the globally Lorentzian spacetime that results from
excising the critical points from a Morse geometry.  We note that a Morse
spacetime is partially ordered by the causality relation, since the Morse
function is a global time function and precludes CTCs. In the discussion
section we return to this construction and suggest a way of extending the
causal structure to the Morse geometry with the degenerate points left in.

A general theorem would run along the following lines: a
Morse spacetime $(M,g)$ is causally continuous {\it iff\/} none of the
excised Morse points $\{p_k\}$ have
index $1$ or $n-1$.  We do not prove the full theorem in this
paper, but examine a specific class of spacetimes defined  on a 
neighbourhood of a single critical point and 
are able to verify the conjecture in these cases. 

In section 2 of this paper, we remind the reader of some 
standard definitions and properties of the causal structure of
Lorentzian spacetimes, in particular the definition of
causal continuity. 
 
Section 3 contains preliminary general results on the causal continuity of
two important classes of spacetimes. First we show that a Morse 
geometry with no degeneracies, which necessarily has
topology $\Sigma \times  [0,1]$ (where 
$\Sigma$ is a closed $n-1$ manifold), is causally continuous. This follows as 
a corollary to a general result on the equivalence of various
causality conditions in the case of a compact cobordism. Another consequence
of this general result is that imposing  
strong causality on the histories of the SOH 
in the case of a compact product cobordism is equivalent to
restricting the sum to non-degenerate Morse metrics. 
This gives us additional confidence that the proposal to 
sum over Morse metrics is a good one.
Second, the general sphere creation/destruction
elementary cobordism, which we refer to as the ``yarmulke''
spacetime, is shown to be causally continuous.  
  
In section 4 we briefly describe the Morse spacetimes that we study in the
rest of the paper.  These are simple ball-neighbourhoods of the critical
points of a Morse function $f$, with the critical point excised, on which
the Morse metric  is constructed from $f$ and a Riemannian metric which is
flat and Cartesian in the  coordinates in which $f$ takes its canonical form.

Section 5 contains a detailed analysis, in two spacetime
dimensions, of the causal structure of these neighbourhood spacetimes for 
index  $\lambda =0$ (the yarmulke) and
$\lambda =1$ (the trousers). These are, up to time reversal, the two basic types of topology
change in two dimensions.  
The general proof in Section 3 shows that the two dimensional
yarmulke is causally continuous.  We verify that the trousers 
neighbourhood geometry is causally discontinuous.

Section 6 contains our main results on the neighbourhood geometries.  We
show that a neighbourhood geometry is causally continuous {\it iff\/} 
its Morse point does not have index 1 or $n-1$. The proof makes extensive
use of the causal structure of the two dimensional yarmulke and trousers
from the previous section.

We summarise our results in the section~7 and comment on further
aspects of this work that are currently under investigation.

\section{Causal continuity}\label{causalcontinuity.sec}

{\it Causal continuity\/} of a spacetime means, roughly, that the volume of
the causal past and future of any point in the spacetime increases or
decreases continuously as the point moves continuously around the
spacetime. Hawking and Sachs~\cite{sachs73}
give six concrete characterisations of causal continuity, three of which
are equivalent in any globally Lorentzian, time-orientable spacetime, while
the equivalence to the remaining three further requires that the spacetime
be distinguishing.  A time orientation is
defined in a spacetime $(M,g)$ by the choice, if possible, of a 
nowhere vanishing
timelike vector field $u$.  A spacetime is called distinguishing if
any two distinct points have different chronological pasts and different
chronological futures.

We take a timelike or null vector $v$ to be future pointing if $g(v,u)<0$ and past
pointing if $g(v,u)>0$.  We define a {\it future directed timelike curve\/} in $M$ to
be a $C^1$ function $\gamma : [0,1] \rightarrow M$ whose tangent vector is
future pointing timelike at $\gamma(t)$ for each $t\in [0,1]$. We also use the
phrase future-directed timelike curve and the symbol $\gamma$ to denote the 
image, $\{x\in M: x =\gamma(t), t\in [0,1]\}$, of such a function. Strictly we should call the function a
``path'', say, and reserve ``curve'' for the image set, but we will ignore this
 distinction for ease of notation, since no ambiguity arises in what follows. Future directed causal curves are defined
similarly, but the future directed tangent vector can be null as well as timelike
and the curves are allowed to degenerate to a single point. Past directed curves
are similarly defined using past pointing tangent vectors.  

We write $x << y$ whenever there is a future directed
timelike curve $\gamma$ with
$\gamma(0)=x$ and $\gamma(1)=y$ and $x <y$ whenever there is a 
future directed causal curve
$\gamma$ with $\gamma(0)=x$ and $\gamma(1)=y$. The chronos relation $I \subset
M\times M$ is defined by $I \equiv \{(x,y): x<<y \}$  while the
causal relation $J\subset M\times M$ is defined by $J \equiv \{(x,y): x<y \}$.
The chronological future and past of a particular point
$x\in M$ are, $I^+(x) \equiv \{y: (x,y) \in I\}$ and
$I^-(x)\equiv \{y: (y,x) \in I\}$, respectively. The
causal future $J^+(x)$ and the causal past $J^-(x)$ of a point are
similarly defined.

It can be shown, using local properties of the lightcone, that the
chronological relation is transitive (i.e., $x<<y , \> y << z 
\Rightarrow x <<z$) and that it is open as a subset of $M\times M$.  The
relation $J$ is transitive and reflexive (i.e., $x<x$)
\cite{kronheimer68}. In simple spacetimes such as Minkowski, $J^{\pm}(x)$
is also closed as a set in $M$, but in general it is not so.  Given a
subset $U$ of the spacetime, $I^+(x,U)$ denotes the set of points in $U$
that can be reached from $x$ along future directed timelike curves totally
contained in $U$. Note that $I^+(x,U)\subset I^+(x)\cap U$, but the
converse is not true in general. $I^-(x,U)$ is similarly
defined. Henceforth the dual or time-reversed definitions and 
statements are understood unless stated otherwise.
 
We note that the Morse spacetimes (\ref{mmetric.eq}) are time-oriented
and distinguishing.  The time orientation is given by the timelike vector
field normal to the level surfaces of the Morse function 
$f$. Distinguishability is guaranteed by the fact that
$f$ is a global time function: if $y$ had the same future set as $x$, then
$y$ would be in $\overline{I^+(x)}$ and it would have to lie in the same
level surface $f^{-1}(a)$; but looking at a convex normal
neighbourhood \cite{penrose72} of $x$, we see that $x$ is the only point in
$\overline{I^+(x)}\cap f^{-1}(a)$.

Thus, all six characterisations of 
causal continuity given in~\cite{sachs73} are equivalent for
Morse geometries. Before we list four of these 
we first give a few more definitions.

The {\it common past\/}, $\downarrow U$ ({\it common future\/}, $\uparrow U$)
of an open set $U$ is the interior of the set of all
points connected to each point in $U$ along a past (future) directed 
timelike curve, i.e., 
\bea 
\downarrow U &\equiv& Int\left(\{x : \; x<< u \quad \forall u\in
U\}\right)\nonumber \\  
\uparrow U & \equiv& Int\left(\{x : \; x>>u \quad \forall u\in U\}\right).  
\eea
We state two properties of the past and future sets are used later. 
If $x$ is a point in the time-oriented spacetime $M$, then: (i)
$J^{\pm}(x)\subset  \overline{I^{\pm}(x)}$ (Proposition
3.9\cite{penrose72}) (ii)  $I^+(x)\subset \uparrow I^-(x)$ and
$I^-(x)\subset \downarrow I^+(x)$ (Proposition 1.1\cite{sachs73}). 

Let $F$ be a function which assigns to each event $x$ in $M$ an open set
$F(x) \subset M$. Then $F$ is said to be {\it outer continuous\/}
if for any $x$
and any compact set $K \subset M - {\overline {F(x)}}$, there exists a 
neighbourhood $U$ of $x$ with $K \subset M - {\overline {F(y)}}$ 
$\forall y \in U$.

A time-orientable distinguishing spacetime, $(M,g)$ is said to be 
{\it causally continuous\/} if it satisfies any of the 
equivalent properties:
\begin{enumerate}

\item  for all events $x$ in the interior of $M$ we have $I^+(x)=\uparrow I^-(x)$ and
$I^-(x)=\downarrow I^+(x)$. 
\item $(M,g)$ is reflecting, i.e., for all events $x$ and $y$ in $M$,
$I^-(x)\subset I^-(y)$ {\it iff\/} $I^+(y)\subset I^+(x)$; 
\item  for all events $x$ and $y$, $x\in \overline{J^+(y)}$ 
{\it iff\/}  $y\in \overline{J^-(x)}$; 
\item for all events $x \in M$, $I^+(x)$ and $I^-(x)$ are outer continuous.
\end{enumerate}

Although the last characterisation might seem to capture better our
intuitive understanding of causal continuity, it is the first one that we
use widely in this paper\footnote{Note that the set identities in condition
1 would trivially fail at the initial and final boundaries of a Morse
spacetime, while the other conditions hold everywhere. In \cite{sachs73}
condition 1 appears without the specification ``the interior of''. There a
spacetime is implicitly assumed to be a genuine manifold without
boundary. With our specification we are ensuring that the conditions above
are still equivalent in the case of a Morse spacetime and preventing the
trivial causal discontinuities at the boundaries.}. For this reason, we
introduce the following point-by-point criterion. A spacetime $(M,g)$ is
{\it causally continuous at point\/} $x$ if $I^+(x)=\uparrow I^-(x)$ and
$I^-(x)={\downarrow I^+(x)}$. Thus $(M,g)$ is causally continuous {\it iff
\/} it is causally continuous at every point of $M$.

We also require the definitions of {\it causality},  
{\it strong causality}, {\it stable causality}, {\it causal simplicity\/}
and {\it globally hyperbolicity}.  
{\it Causality\/} holds in a subset $S$ of $M$ if  there are no
causal loops based at points in $S$.
{\it Strong causality\/} holds in a subset $S$ of $M$ if
for
every point $s\in S$ any neighbourhood $U$ of $s$ contains another
neighbourhood $V$ of $s$ that no causal curve intersects more than once.
A spacetime $(M,g)$ is said to be {\it stably causal\/} 
if there is a metric $g'$, whose lightcones are strictly broader than 
those of $g$ and for which the spacetime $(M,g')$ is causal.
A spacetime is stably causal if and only if it admits a global time
function, i.e., a function whose gradient is everywhere timelike.
A spacetime is said to be {\it causally simple\/} if
$J^+(q)$ and $J^-(q)$ are closed for every point $q$. This is equivalent
to the conditions $J^+(q)={\overline{I^+(q)}}$ and $J^-(q)={\overline{I^-(q)}}$.
A spacetime is said to be {\it globally hyperbolic\/} if it contains
a spacelike hypersurface which every inextendible 
causal curve in the spacetime
intersects exactly once.
There is a standard sequence of implications amongst these 
causality conditions~\cite{sachs73}: 
global hyperbolicity of a spacetime
$\Rightarrow$  causal simplicity $\Rightarrow$  causal continuity  
$\Rightarrow$ stable causality $\Rightarrow$  strong causality $\Rightarrow$
causality.

Finally, we recall a result that we use in our proofs: the
Causal Curve Limit Theorem (CCLT) \cite{penrose72}, which states that if
$K$ is the set of all points in $M$ where strong causality holds and
$C\subset K$ is compact, then for any closed subsets $A$, $B$ of $C$ the
space $\mathcal{C}_C(A,B)$ of causal curves in $C$ from $A$ to $B$ is
compact.  Strictly speaking, this holds only if one uses a weaker
definition of causal curve than the one we are using since the limit 
curve $\gamma$ to which a sequence of $C^1$ causal curves converges 
is not
in general $C^1$.  The existence of the limit curve nevertheless 
ensures the existence of some
$C^1$ causal curve between the endpoints of the limit curve, and this
subtlety is ignorable for our purposes.

\section{Causal continuity of non-degenerate and 
yarmulke spacetimes}
\label{ccproduct.sec}

By ``non-degenerate spacetime'' we mean a compact Morse spacetime 
$(M,g)$, {\it i.e.} a Morse geometry with no degeneracies.
A yarmulke spacetime is one arising from a Morse geometry  
in an $n$-dimensional elementary cobordism of index
$\lambda =0\;(n)$, whose initial (final) boundary is empty and 
final (initial) boundary is $S^{n-1}$.  

In what follows all spacetimes are assumed to be time-orientable and
distinguishing. We start by establishing a general result on the behavior of
compact, globally Lorentzian spacetimes:

\begin{proposition}\label{compactspaces.prop}
For a compact spacetime $(M,g)$, the following properties are equivalent:
\begin{enumerate}
\item It is causally simple.
\item It is causally continuous.
\item It is stably causal. 
\item It is strongly causal.
\end{enumerate}
\end{proposition}

\noindent {\bf Proof:} Since each item implies the following we only need to
prove that the last item implies the first. (Indeed, one can readily check that 
(1)~$\Rightarrow$~(2). That (2)~$\Rightarrow$~(3) is the content 
of proposition 2.3 ~\cite{sachs73} and that (3)~$\Rightarrow$~(4) is shown in for
example \cite{hawking73}.) 

So suppose that strong causality holds on $(M,g)$.
For any spacetime, $J^+(q)\subset {\overline{ I^+(q)}}$. 
Let $p \in {\overline {I^+(q)}}$, and consider a 
sequence of points $p_k \in{I^+}(q)$,  which
converges to $p$. 
There must be a sequence of future-directed timelike curves, 
$\gamma_k$ from $q$ 
to $p_k$.  Since $(M,g)$ is strongly causal, we can use the CCLT 
by taking
$C=M$, $A = \{q\}$, $B = \{p_k: k = 1,2,\dots\} \cup \{p\}$, so that there
is a causal limit curve $\gamma$ from $q$  to $p$. Thus,  $p \in
{J^+}(q)$, or ${\overline{ I^+(q)}}\subset J^+(q)$ which implies that
$J^+(q)={\overline{ I^+(q)}}$. $J^-(q)=\overline{I^-(q)}$ is proved similarly.
So $(M,g)$ is causally simple $\Box$.

This proposition provides us with a proof of the causal continuity
of non-degenerate Morse spacetimes $(M,g)$, since they are compact and 
possess a global time function:

\begin{corollary}
A non-degenerate Morse spacetime $(M,g)$ is causally continuous.
\end{corollary} 

\begin{proposition}\label{compactspaces2.prop}
Let $(M,g)$ be a compact spacetime 
with boundary $\partial M = V_0 \coprod V_1$ such that 
$V_0$ and $V_1$ are closed $n-1$ manifolds which are 
the initial and final spacelike boundaries of $M$ respectively.
Then the following properties of $(M,g)$ are equivalent:
\begin{enumerate}
\item It is a non-degenerate Morse spacetime.
\item It is stably causal.
\item It is globally hyperbolic.
\end{enumerate}
\end{proposition}

This result, as stated, does not ask that $V_0$ and $V_1$ be non-empty or
have the same topology but the known causality violations in those cases
makes the result relevant only to spacetimes with these properties. Its
significance is that for this class of spacetimes global hyperbolicity can
be added to the equivalent conditions in proposition
\ref{compactspaces.prop} and that any Lorentz metric in a strongly causal
product cobordism can be written as a Morse metric.


{\bf Proof:} 

a) 1 $\Rightarrow$ 2. For any Morse spacetime $(M,g)$ with its critical 
points excised the Morse function is a global time function and hence~\cite{hawking73} 
the spacetime is stably causal. 

b) 2 $\Rightarrow$ 1. Suppose that $(M,g)$ is stably causal with time 
function~$f$. Let $b^2=-g^{\mu\nu}\partial_\mu f \partial_\nu f $. Define the
positive definite metric $h$ by
\be
h_{\mu\nu} \equiv g_{\mu\nu} + (1+ 1/b^2)\partial_\mu f \partial_\nu f.
\ee
(That $h$ is positive definite may be checked by using the basis
$\{T^\mu, S^\mu_i\}$, where $T^\mu=g^{\mu\nu}\partial_\nu f$  
is timelike with respect to $g$ and the $S^\mu_i$ are chosen to
be spacelike and orthogonal to each other and to $T^\mu$ (with respect 
to $g$). Then $\{T^\mu, S^\mu_i\}$ forms an orthogonal basis with respect 
to $h$, and all the vectors have positive norm.)
Since $h^{\mu \nu} \pd_\mu f \pd_\nu f=1$, we see that we may 
invert the above expression and express $g$ as a Morse metric of the 
form~(\ref{mmetric.eq}). 

It is possible to choose~$f$ such that $f^{-1}(0)
= V_0$ and $f^{-1}(1) = V_1$. 
The function 
$\hat f(p) \equiv Volume(\overline{I^-(p)})$ is a time
function on~$(M,g)$.  Let $T^\mu \equiv g^{\mu\nu} \partial_\nu \hat f$.
For any point $p\in M$, let $q_p\in V_1$ be the future endpoint
of the integral curve of $T^\mu$ through~$p$. Define $v(p)\equiv
Volume(\overline{I^-(q_p)})$. Then $f(p) \equiv (1/v(p))\hat f(p)$ 
is a time function on $M$ with $f^{-1}(0)=V_0$ and $f^{-1}(1)=V_1$.

c) 2 $\Rightarrow$ 3. Suppose that $(M,g)$ is stably causal, with time
function $f$.  Let $S$ be any $f=\hbox{constant}$ surface.  Let $p$
lie on some later time surface, and suppose that a past inextendible
causal curve from $p$, say $\lambda$, does not intersect~$S$.  Then
$\lambda$ is trapped in the region of $M$ between the two
constant-time surfaces mentioned above. Since this region is compact
\cite{milnor65} $\lambda$ must accumulate at some point~$q$ and
therefore it intersects the constant-time surface through~$q$ more
than once.  This is not possible. $\Box$

Proposition \ref{compactspaces2.prop} shows that the restriction to
Morse metrics in a Lorentzian compact spacetime with initial and final
spacelike boundaries is a reasonable one, since that restriction is equivalent to 
stable causality which is in turn equivalent to strong causality.   

We now  prove causal continuity for yarmulke spacetimes in $n$
dimensions. The Morse function has a single critical point $p$. 
Suppose that $f:M\to [0,1]$ is the Morse function and
that the boundary of $M$ is a final boundary. Then we must have $f(p)=0$.
Excising $p$ from our manifold, the associated Morse spacetime has topology 
$S^{n-1}\times (0, 1]$. 
\begin{lemma}
\label{yarmulke.lemma}
The yarmulke spacetime $(M,g)$, with topology $S^{n-1} \times (0,1]$, is 
causally continuous.
\end{lemma}

\noindent
{\bf Proof:} Using arguments similar to those in used in the above
proposition,  we see that since the spacetime is stably causal, and
therefore strongly causal, the surfaces  of constant $f$ are
Cauchy surfaces. (The full spacetime is not compact here, but the region
between any two level surfaces of $f$ is.) Thus the spacetime is 
globally hyperbolic and hence causally continuous. $\Box$


\section{The neighbourhood of a  Morse point}
\label{csnbhoodMp.sec}

In this section we define the neighbourhood Morse spacetimes for 
which we verify the Borde-Sorkin conjecture. Consider an open ball, 
$D_\epsilon$, of radius $\epsilon$
in $\mathR^n$ centred on the point $p$.  Let  
$\{x^i, y^j: i= 1, \dots \lambda, j = 1, \dots
n - \lambda\}$, be local coordinates
with $x^i(p)=y^j(p)=0$, $\sum_i (x^i)^2 + \sum_j (y^j)^2 < \epsilon^2$
and  in which the Morse function $f$ takes 
the following canonical form (Morse lemma \cite{milnor65}):
\be 
f = f(p) - \sum_i (x^i)^2 + \sum_j (y^j)^2 \ .
\ee
The spacetime manifold
we are concerned with is then $\N \equiv D_\epsilon - \{p\}$.
Henceforth we consider all set closures, {\it etc.}, to be taken 
in the manifold $\N$ except when explicitly stated otherwise. 
We frequently refer to $p$ as though it is in the
spacetime: for example we refer to sets of curves that are  ``bounded
away from $p$'' the meaning of which should be clear.  

We define the polar coordinates, $(\rho, \Theta, r,
\Phi)$, where $\rho ^2=\sum_1^{\lambda}x_i^2 $, $\; r^2
=\sum_1^{n-\lambda}y_j^2$, and the collective coordinates $\Theta$ and
$\Phi$ stand for the angles $\theta_i, \quad i=1\cdots \lambda\! -\! 1$ and
$\phi_j, \quad j=1 \cdots n\!  -\!\lambda\! -\! 1$ that coordinatise the
$(\lambda -1)$-sphere and the $(n-\lambda -1)$-sphere, respectively.  When
$\lambda = 1$ the sphere $S^{0}$ is disconnected and our convention for
such cases is to adopt a single discrete ``coordinate'' $\theta_0$
with two possible values: $0$ and $\pi$. Then the transformation between
$x^1$ and $(\rho,\theta_0)$ is $x^1=\rho \cos\theta_0$. Similarly, when
$\lambda = n\!  -\!  1$, the disconnected sphere would be parameterised by
the single discrete angle $\phi_0$ and $y^1=r\cos \phi_0$. 
We will often use the notation whereby the coordinates of a point $q$ are
referred to as $(x^i_q, y^j_q)$ or $(\rho_q, \Theta_q, r_q, \Phi_q)$ 
except that the discrete angles are written $\theta_0(q)$ or $\phi_0(q)$
for ease.

The particular Morse metrics that we study in this
paper are those for which the auxiliary 
Riemannian metric $h$ is the flat Cartesian
metric in the coordinates $\{x^i, y^j\}$ introduced above. In that case
we find, transforming to polar coordinates, the metric  (\ref{mmetric.eq}) 
becomes 
\begin{eqnarray}  
ds^2 & = & 4\{ (\rho^2 + r^2)(\rho ^2 d\Theta_{\lambda - 1}^2 + r^2
  d\Phi_{n-\lambda -1}^2) + (r^2-(\zeta -1)\rho^2)d\rho^2 \nonumber \\  
& &  +(\rho^2-(\zeta -1)r^2)dr^2 + 2 \zeta \rho rd\rho dr\}. 
\label{sphmetric.eq}
\end{eqnarray}
In these coordinates $ds^2$ is seen to be a warped product metric  
\footnote{A metric $g(x^a, y^A)$ is a warped product metric if its interval
splits as $ds^2 = g_{ab}({\vec x}, {\vec y})dx^adx^b + g_{AB}({\vec y})dy^A
dy^B$.} since it decomposes into a radial and angular part, i.e., $ds^2 =
ds^2_R +ds^2_A$, with the radial coordinates $(r,\rho)$ warping the angular
part.  An important property of such a warped product form,
is that the geodesics of the metric $ds_R^2$ are also geodesics of the full
metric. 
We note also, that $ds_A^2$ is never negative, so that  for any timelike
(causal) 
curve $\gamma(t)=(\rho(t), \Theta (t), r(t),\Phi(t))$,
the related curve  $\gamma'(t)=(\rho(t), \Theta_a , r(t),\Phi_b)$ at 
any fixed angles $\Theta_a$ and $\Phi_b$, is also timelike (causal).
The Morse function $f(r, \rho) = f(p) -\rho^2 + r^2$ must increase along
future directed timelike curves so that the future time direction is,
roughly speaking, decreasing $\rho$ and increasing $r$. 

We call $(\N,g)$, where $g$ is given by
(\ref{sphmetric.eq}),  the {\it neighbourhood geometry \/} of type $(\lambda,
n-\lambda)$.

\section{Causal structure for $n=2$}
\label{2d.sec}

We study in detail the two elementary neighbourhood geometries ---up to
time  reversal--- in $n=2$,   namely the trousers ($\lambda=1$)  and the 
yarmulke ($\lambda =0 $.)  The causal structure of these
two cases turns out to be crucial in studying the general $n$-dimensional
neighbourhood geometries. (A study of the causal structure of the $1+1$
trousers for a particular choice of flat metric was first carried out by
\cite{dray}.)   
\begin{figure}[ht]
\centering
\rotatebox{270}{\resizebox{!}{4in}{\includegraphics{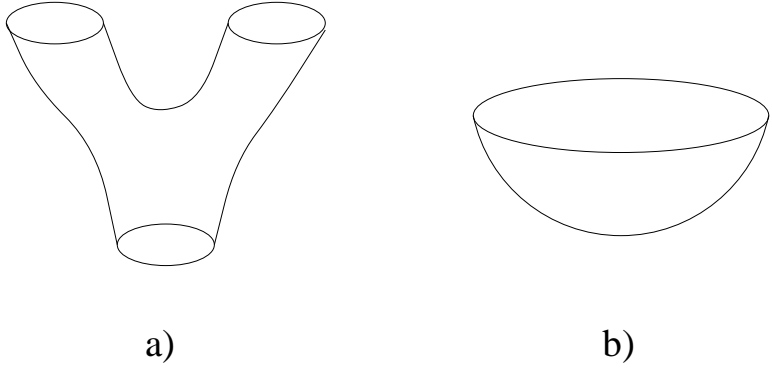}}}
\caption{{\small Global topology of the trousers and yarmulke
cobordisms. In the trousers $\N$ is to be regarded as a little region around 
the saddle point, while in the yarmulke $\N$ has the same
topology as the whole cobordism, which is just a disc.}\label{trouyar.fig}}
\end{figure}

\subsection{Causal structure in the trousers}\label{trousers.subsec}

\begin{figure}[ht]
\centering
\rotatebox{270}{\resizebox{!}{2.5in}{\includegraphics{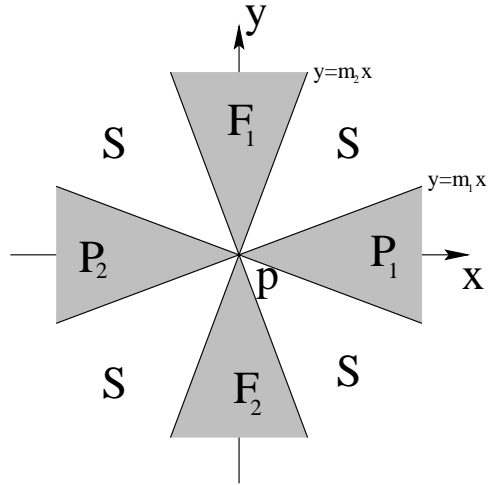}}}
\caption{{\small The trousers. Partition of the disc by radial geodesics. The
timelike geodesics are in the shaded regions, future outward
about the y-axis and future inward about the x-axis. The radial
geodesics outside the shaded regions are spacelike.}
\label{tpart.fig}}
\end{figure}

The neighbourhood Morse metric for the trousers is  
\be 
\label{tinterv.eq}
ds^2  =  4\{ (y^2-(\zeta -1)x^2)dx^2 +(x^2-(\zeta -1)y^2)dy^2 + 2 \zeta
xy dx dy\}. 
\ee 
Any radial line with endpoint at the origin is a geodesic so the norm of its 
tangent vector has the same sign all along. Thus the disc can be
partitioned into  sectors that are loosely speaking, either future
time-like, past time-like or space-like related to the origin
(see figure \ref{tpart.fig}).   
Indeed by substituting for $y=mx$ in the interval  one obtains:
\be
ds^2 = -4((\zeta -1)m^4-2(\zeta +1)m^2+(\zeta- 1))x^2dx^2\quad  \left\{\begin{array}{ll}
                                         >0 \;  &\mbox{if}\quad (m_1)^2<m^2 <(m_2)^2 \\
                                   \\
                                   \leq 0  &\mbox{otherwise}
                                         \end{array}\right.
\label{m1m2.eq}
\ee
where the gradients $m_1={\sqrt{\zeta}-1\over \sqrt {\zeta -1}}< 1$ and 
$m_2 ={\sqrt{\zeta}+1\over {\sqrt {\zeta -1}}} = m_1^{-1}> 1$  
mark the transition
between the spacelike and timelike character of the radii, with  
the lines $y=\pm m_1x$ and  $y=\pm m_2x$ being null and
separating the sectors. 

We define sets $\mathcal{P}_1, 
\mathcal{P}_2,
\mathcal{F}_1$ and $\mathcal{F}_2,$  via
\bea
\mathcal{P}_1 &\equiv& \{({x}, {y})\in \N: |y|<m_1 |x|\ {\rm{ and}}\ x>0\}\nonumber\\ 
\mathcal{P}_2 &\equiv& \{({x}, {y})\in \N: |y|<m_1 |x|\ {\rm{ and}}\ x<0\}\nonumber\\ 
\mathcal{F}_1 &\equiv& \{({x}, {y})\in \N: |y|>m_2 |x|\ {\rm{ and}}\ y>0\}\nonumber\\ 
\mathcal{F}_2 &\equiv& \{({x}, {y})\in \N: |y|>m_2 |x|\ {\rm{ and}}\ y<0\}\\ 
\label{fandp}
\eea
Also $\mathcal{P} \equiv \mathcal{P}_1 \cup \mathcal{P}_2$, and
$\mathcal{F} \equiv \mathcal{F}_1 \cup \mathcal{F}_2$. We define
$\partial\mathcal{P}$ and $\partial\mathcal{F}$:    
\bea
\partial\mathcal{P} &\equiv& \{(x,y)\in \N: y = m_1 x \ {\rm {or}}\ y = - m_1 x\}\nonumber\\ 
\partial\mathcal{F} &\equiv& \{(x,y)\in \N: y = m_2 x \ {\rm{ or}}\ y = -m_2 x\}
\label{bounfandp}
\eea
and $\mathcal{S} \equiv \N - (\mathcal{P}\cup\mathcal{F}\cup \partial\mathcal{P} 
\cup \partial\mathcal{F} )$.
We see that $\mathcal{F}$ is what we'd expect for the 
chronological future of the Morse point, $p$,  $\partial\mathcal{F}$  
is what we might want to call the future lightcone of $p$ and similarly for the
past; $S$ is the ``elsewhere'' of $p$.  The status of these sets is
discussed further in a later subsection. 

To summarise, through all the points in the shaded regions of figure
(\ref{tpart.fig}) there passes a radial timelike geodesic; $S$ denotes the
points outside the shaded regions through which the radial geodesics are
spacelike, while the boundary radial geodesics, with gradient $\pm m_1$ and
$\pm m_2$ are null.

We return to the general null geodesics which give us the lightcones
for an arbitrary point. This is a generalisation to arbitrary $\zeta$ of
the analysis of \cite{surya97}.  From equation 
(\ref{tinterv.eq}) one obtains an implicit expression for the null curves: 
\bea
\label{veeplus}  \sqrt{\zeta -1}(x^2 - y^2) &=&  2xy + c_{+ }\\
\label{veeminus} {\rm or} \qquad \sqrt{\zeta -1}(x^2 - y^2) &=& - 2xy + c_{-}\ ,
\eea
where $c_{\pm}$ are constants.  To get a clearer idea of what these curves are 
consider a rotated coordinate system 
$(x',y')$ with the $x'$ axis being the line $y = m_1 x$ at an angle of
$\alpha = \tan^{-1}m_1$ with the $x$ axis:
\be
\label{xprime.eq}
\left(\begin{array}{c} 
	x' \\ y' \end{array}\right) =
\left( \begin{array}{rr} 
	\cos\alpha & \sin\alpha\\ -\sin\alpha & \cos\alpha \end{array}\right)
\left( \begin{array}{c} 
	x \\ y \end{array}\right)
\ee
Then (\ref{veeplus}) takes the form 
\be
\label{hyperb1.eq}
                             x'y'=\frac{-c_+}{2\sqrt\zeta} 
\ee
which shows that they are hyperbolae with $y=m_1x$ and $y=-m_2x$ as asymptotes.
Rotating to coordinates $(x'',y'')$ by $-\alpha$ instead, (\ref{veeminus}) becomes
\be
\label{hyperb2.eq}      x''y''=\frac{c_-}{2\sqrt\zeta} 
\ee
so these are hyperbolae with $y = -m_1x$ and $y = m_2 x$ as asymptotes.

Through every point $q$ in $\N$ there passes a curve which satisfies
eq. (\ref{hyperb1.eq}) with a particular $c_+$, we call it $\upsilon_q^+$
and another curve which satisfies eq. (\ref{hyperb2.eq}), we call this
$\upsilon_q^-$.  As we see shortly, these null geodesics through $q$
suffice to bound its past and future provided $q$ lies in $S$, but not
otherwise. To determine systematically the
chronological pasts and futures of all points in the disc, we start by
noting that in the rotated coordinate systems,
\be
\label{rota.eq}
\left(\begin{array}{c} 
	\tilde {x}\\ \tilde{y}\end{array}\right) =
\left( \begin{array}{rr} 
	\cos\psi & \sin\psi\\ -\sin\psi & \cos\psi \end{array}\right)
\left( \begin{array}{c} 
	x \\ y \end{array}\right)
\ee
the hyperbolae given by $\tilde{x}\tilde{y}=$constant are timelike
if and only if $-\alpha <\psi <\alpha$ or ${\pi \over 2} -\alpha <\psi<{\pi \over 2}+ 
\alpha$, that is when the $\tilde{x}$ and $\tilde{y}$ axis fall in the interior
of the shaded regions in figure \ref{tpart.fig}. 
We will use segments of such timelike hyperbolae to determine the
chronological relation.  By symmetry  and time-reversal invariance, we need
consider only three representative points in the upper right quadrant: (i)
$q\in \mathcal{S}$,  
(ii) $q\in \mathcal{F}_1$ and (iii) $q\in \partial\mathcal{F}_1$.

\begin{figure}[ht]
\centering
\rotatebox{270}{\resizebox{!}{3in}{\includegraphics{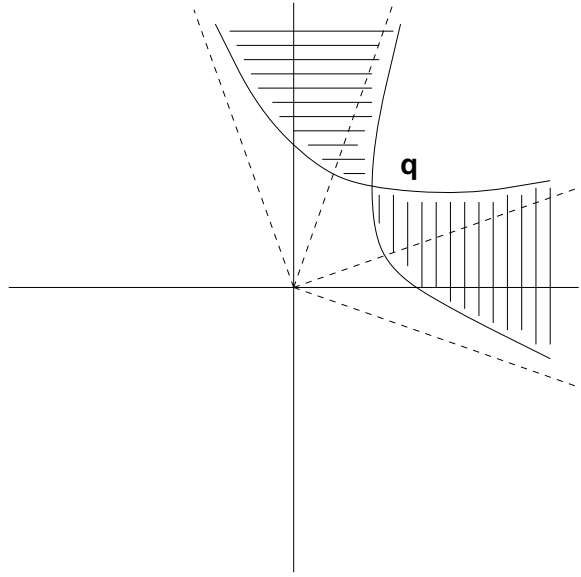}}}
\caption{{\small The trousers. 
$q\in S$ and $I^+(q)$ ($I^-(q)$) is the horizontally  
(vertically) shaded region.
} \label{tS.fig}}
\end{figure}

	(i) For $q\in S$, we claim that $I^+(q)$ is the interior of
the horizontally shaded region in figure \ref{tS.fig}, bounded by the
two null hyperbolae through $q$, $\upsilon_q^+$ and $\upsilon_q^-$ and
that $I^-(q)$ is the interior of the vertically shaded region between
the same hyperbolae. Those regions are contained in $I^+(q)$ and
$I^-(q)$, because they are swept out by the timelike hyperbolae
through $q$ between $\upsilon_q^+$ and $\upsilon_q^-$.  To see that
such regions exhaust all of $I^+(q)$ and $I^-(q)$ is also
straightforward. Any future directed timelike curve from $q$ must
begin by heading into the horizontally shaded region. If there was one
such curve ending at a point $s$ outside the region, it would have to
intersect one of the bounding hyperbolae at some point $q'$, but the
tangent vector there could not point out of the region and be both
timelike and future directed according to the local lightcone at
$q'$. Similarly for $I^-(q)$.

\begin{figure}[ht]
\centering
\rotatebox{270}{\resizebox{!}{4in}{\includegraphics{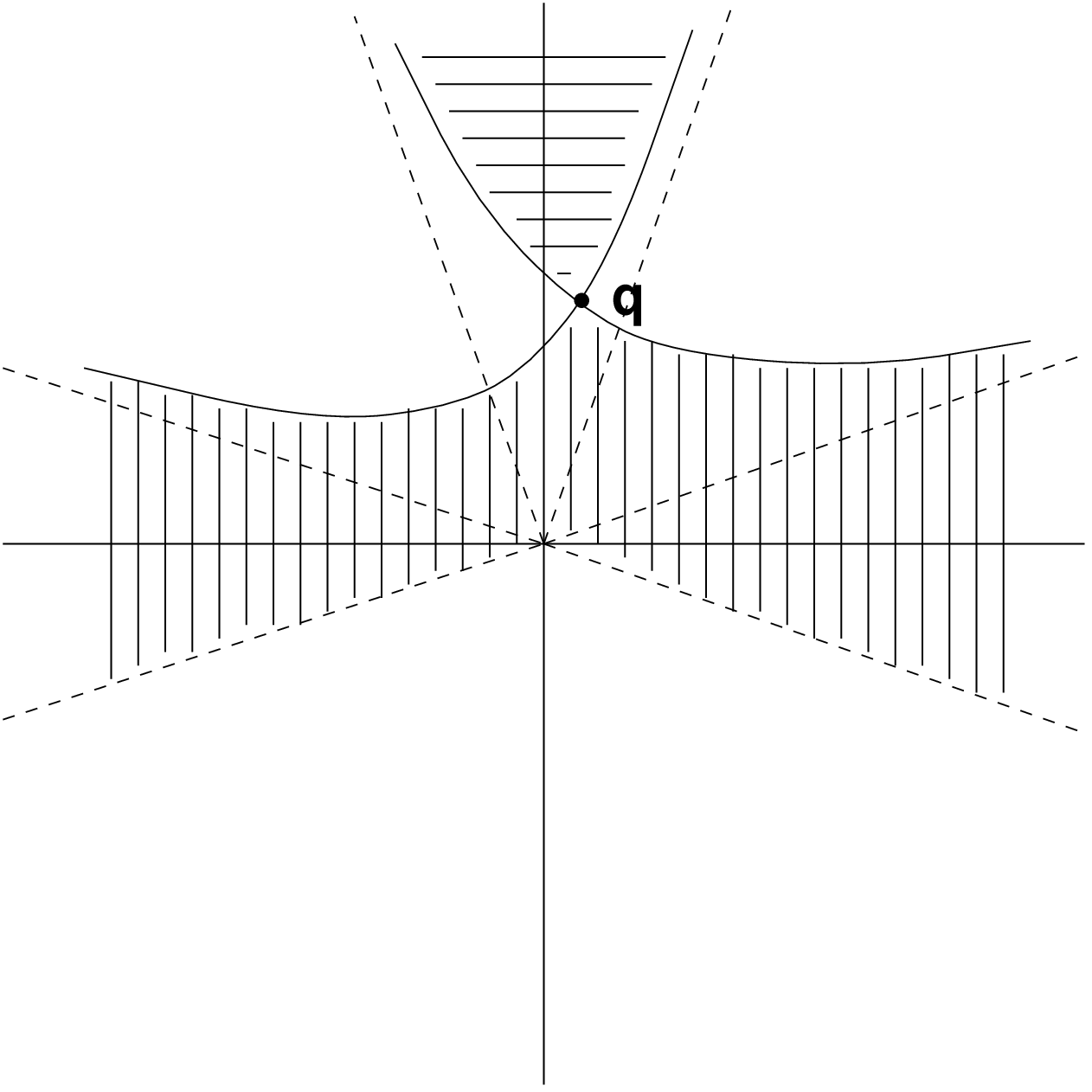}}}
\caption{{\small The trousers.
$q\in \FU$ and $I^+(q)$ ($I^-(q)$) is the horizontally  
(vertically) shaded region.}
\label{tF.fig}}
\end{figure}

	(ii) For $q\in \mathcal{F}_1$, $I^+(q)$ is the interior 
of the horizontally
shaded region shown in figure \ref{tF.fig}
bounded by the two null hyperbolae through $q$, by the same
arguments as in case (i), and 
we claim that $I^-(q)$ is the interior of the vertically shaded region
bounded by the hyperbolae and by lines $y=m_1x$  and $y=-m_1x $ with
$y < 0$. 

\begin{figure}[ht]
\centering
\rotatebox{270}{\resizebox{!}{4in}{\includegraphics{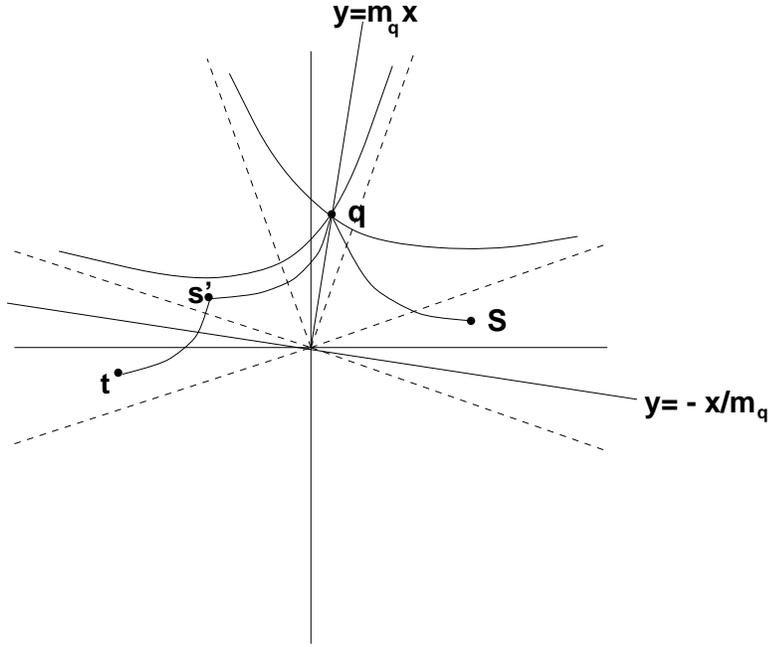}}}
\caption{{\small A point $s\in I^-(q)$ for which $y_s> -x_s/m_q$ 
can be joined to $q$ with a single arc of timelike 
hyperbola. A point $t\in I^-(q)$ with $y_t< - x_t/m_q$ 
requires two such arcs to reach $q$.}
\label{zones.fig}}
\end{figure}

To show that this region is indeed contained in  $I^-(q)$ we find
sequences of timelike hyperbola segments from all points 
in the region to $q$. Let us first divide the region in two zones. 
If $y_q=m_q x_q$, zone 1 consists of points
with $y>-\frac{1}{m_q}x$ and zone 2 of points with $y\le-\frac{1}{m_q}x$
(see figure \ref{zones.fig}).

	A point $s$ in zone 1 can be joined to $q$ by a
single arc of timelike hyperbola whose asymptotes are the 
$\tilde x$ and $\tilde y$ axes defined by (\ref{rota.eq}) with 
$-{1\over m_q}<\tan\psi < m_1$, except when $y_s = m_q x_s$ in which case the 
timelike curve to $q$ is $y = m_q x$.

	For a point $t$ in zone 2 two hyperbolic
arcs are needed: the first one takes it into a point $s'$ in zone
1 via a hyperbola with asymptotes given by (\ref{rota.eq}) with 
$-m_1< \tan\psi < m_t$ where  $m_t = y_t/x_t$. Then
$s'$ can be connected to $q$ as before.

The argument that these regions comprise all of $I^\pm(q)$ is as in (i).

\begin{figure}[ht]
\centering
\rotatebox{270}{\resizebox{!}{4in}{\includegraphics{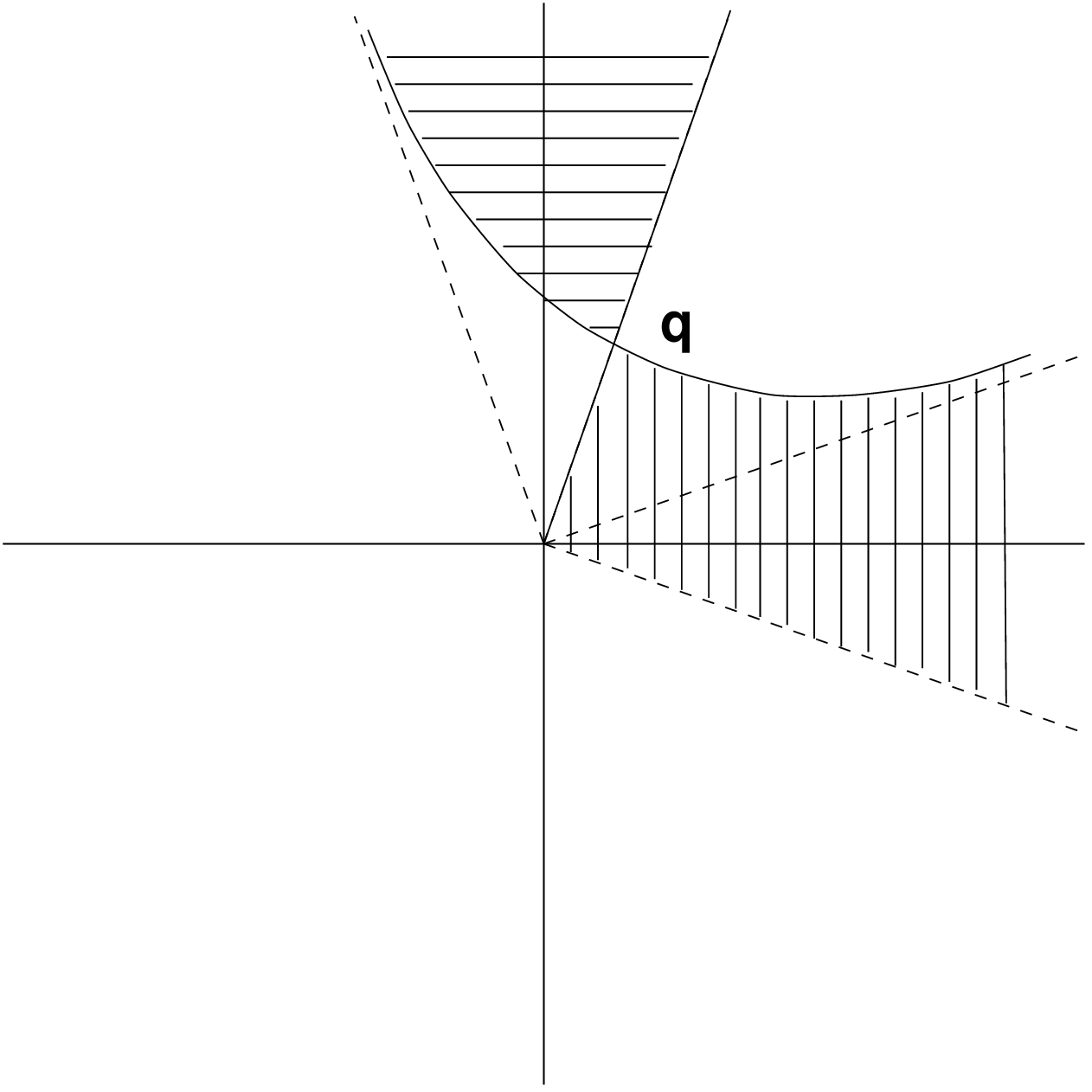}}}
\caption{{\small The trousers. $q\in \partial \FU$ and  
$I^+(q)$ ($I^-(q)$) is the horizontally
(vertically) shaded region.
}
\label{tdF.fig}}
\end{figure}

	(iii) For $q\in \partial\mathcal{F}_1$, similar arguments show that 
$I^+(q)$
is the interior of the horizontally shaded region
bounded by the null hyperbola $x'y' = x'_q y'_q$
and the null line $y = m_2 x$. $I^-(q)$ is the interior of
the vertically shaded region bounded by 
the hyperbola, $y=m_2 x$ and $y = -m_1 x$ 
for $x> 0$ (see figure \ref{tdF.fig}).
Note that $I^-(q)$ does not contain any point with $x<0$. 

Using $J^{\pm}(q) \subset \overline{I^{\pm}(q)}$ the causal pasts and
futures of these representative points are easy to find.  Since all the
points in a null geodesic through $q$ are in its causal past or future, we
just need to decide whether the additional straight lines bounding the
$I^{\pm}$ of points type (ii) and (iii) are in $J^{\pm}$. 
For definiteness, consider $q\in \mathcal{F}_1$ and a point $s$ on 
$y=-m_1x$, $y<0$. Now $q$ is not in $\overline{I^+(s)}$, so it doesn't
belong to $J^+(s)$ either: there is no causal curve from $s$ to $q$. It
follows that the lines that bound $I^-(q)$ in the lower hemiplane are not
in $J^-(q)$. Neither is the line 
$y=-m_1 x$ contained in $J^-(q)$ when $q\in \partial\FU_1$.
Summarising, for points $q\in S$ we have $J^{\pm}(q)=
\overline{I^{\pm}(q)}$, otherwise 
the causal sets $J^{\pm}(q)$ are not closed.

This
completes our analysis of the causal structure around the crotch
singularity in the $1+1$ trousers, which we use extensively later. For
the moment it allows us to establish the following.

\begin{lemma}\label{tcd.lemma} The neighbourhood
geometry $(\N,g)$ of type $(1,1)$
is a causally discontinuous spacetime.
\end{lemma}

\noindent
{\bf Proof:} Let $q\in \partial\mathcal{F}_1$ with $x_q > 0$, then $\cpa \neq \pa$ since
any $s$ on the negative $x$-axis satisfies $s\in \cpa$ but $s\notin \pa. \Box $

\begin{figure}[ht]
\centering
\rotatebox{270}{\resizebox{!}{4in}{\includegraphics{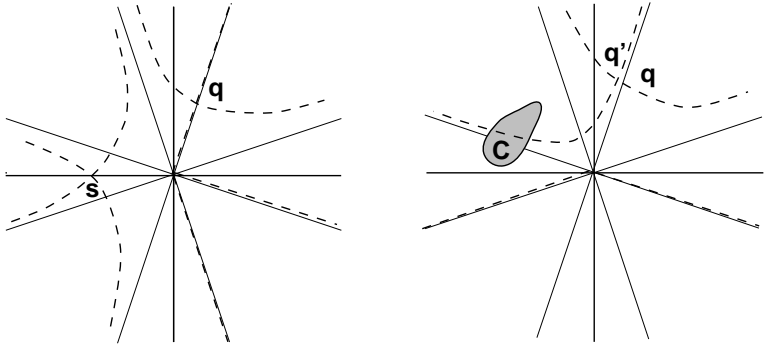}}}
\caption{{\small On the left, the futures and pasts of
a pair of points, $(s,q)$ are shown. 
This pair violates the reflecting property, 
since $I^+(q)\subset I^+(s)$ but
$I^-(s) \not\subset I^-(q)$. Also $q\in \overline{J^+(s)}$ but
$s\protect\notin\overline{J^-(q)}$. On the right the compact set
$C\subset M-\overline{I^-(q)}$ intersects
$\overline{I^-(q')}$ for a point $q'\in \FU_1$ which can be taken arbitrarily
close to $q$.}\label{cdt.fig}}
\end{figure}

In figure \ref{cdt.fig} we illustrate the failure of causal continuity in
each of the remaining three 
equivalent definitions given in section
(\ref{causalcontinuity.sec}), in the hope to give the reader an intuition for their
meaning. 

\subsection{Causal structure in the yarmulke
}\label{yarmulke.subsec} 

We now undertake a similar analysis for the neighbourhood geometry $(0,2)$. 
The Morse metric on the punctured disc is
\be 
ds^2 = 4(-(\zeta -1)r^2dr^2 + r ^4d\phi^2)\ . 
\ee  
Because of the $U(1)$ symmetry
of this  metric,  there is only one class of points.
Since  $f(r) = r^2$ is the time function, timelike and null tangent
vectors are past pointing if their radial component is inwards and future
pointing if it is outwards. As before, in two dimensions 
the geodesic equation is not really
needed to find the null geodesics through a point; solving for a tangent
vector with vanishing norm suffices. The solutions are  the null
``spiraling'' curves $\sigma_0^{\pm}$ given by, 
\be
\label{rphi.eq}
r (\phi) = r_0e^{{\pm\phi \over \sqrt{\zeta-1}}}
\ee
Again these spirals can be shown to be consistent with the geodesic
equations. For geometries with $\zeta \rightarrow 1$, the
lightcones are squeezed onto the radial direction, while for $\zeta
\rightarrow \infty$, the lightcones widen to become circles.

The future direction along the spirals $\sigma_ q^+$ and $\sigma _q^-$
through a point $q$ corresponds to an increase in the radial coordinate,
which, along $\sigma_q^+$  is achieved by increasing $\phi $  and along
$\sigma_q^-$ by decreasing $\phi$. The radial straight lines $\phi=$
constant also satisfy the geodesic equations. These
geodesics are timelike, since $d\phi =0$ along them. 

\begin{figure}[ht]
\centering
\rotatebox{270}{\resizebox{!}{4in}{\includegraphics{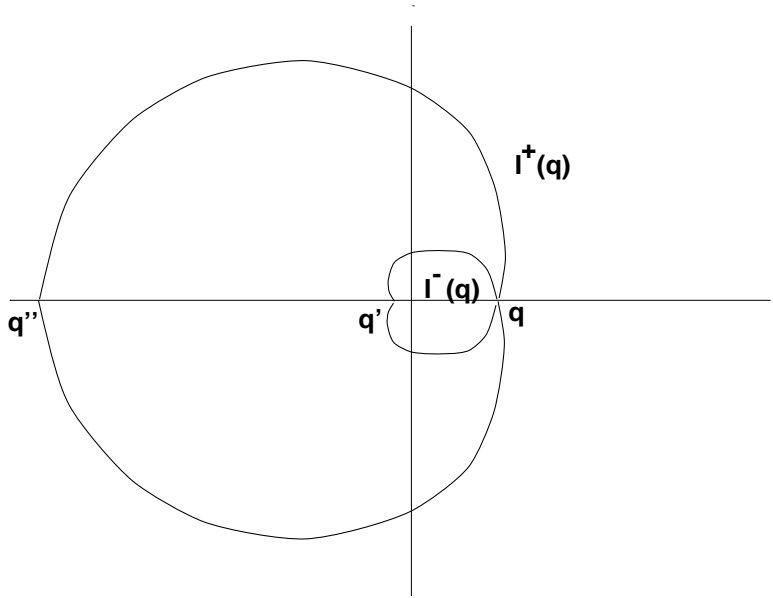}}}
\caption{{\small The yarmulke. 
Segments of the null geodesics through $q=(a,0)$
that bound $q$'s past and future. Their first intersection in the
past occurs at the point $q'=(ae^{-\pi}, \pi)$ and their first intersection
in the future occurs at $q'' = (ae^{+\pi}, \pi)$. 
}\label{ysets.fig}}
\end{figure}

Forgetting for the moment that we are confined to a disc of finite radius,
the two spirals through a point $q$ converge again at both an earlier
$q'$ and at a later $q''$, beyond which these null geodesics no longer bound 
the past or future of $q$ (see figure \ref{ysets.fig}).  Let $L_q$ be the interior of the little
heart-shaped region bounded 
by the two past directed null spirals from $q$ to $q'$ and $B_q$ the big 
heart-shaped region bounded by the 
future directed null spirals from $q$ to $q''$.  
Then 
 (i) $\pa=L_q$ and
(ii) $\fu=\N -B_q$ 

By the symmetry of this metric, all points are
equivalent so that our task reduces to considering the
representative point $q=(r_q,0)$. We only give the argument for (i)
since (ii) is similar.
Let $s=(r_s,\phi_s) \in L_q$.  By the symmetry of $L_q$, we can assume
without loss of generality that $0\le \phi_s \le \pi$.  Thus,
$r_s<r_qe^{{-\phi_s}\over \ze}$, where $\ze= \sqrt{\zeta-1}$. In particular for points $s$ such that $\phi_s=0$,
the radius $\phi =0$ is a timelike curve from $s$ to $q$ and therefore they
belong to $\pa$. If  $\phi_s\neq 0$,  consider the curve 
$\gamma$ from $s$ to $q$ given by
\be
r(\phi) = u(\phi) e^{{-\phi}\over \ze}
\ee
where the function $u(\phi) $,  
\be
u(\phi) =r_q\frac{\phi_s-\phi}{\phi_s} + r_se^{{\phi_s}\over
\ze}\frac{\phi}{\phi_s} 
\ee
decreases smoothly from $r_q$ to $r_se^{{\phi_s}\over \ze}$. Then
\bea
\frac{dr}{d\phi} &=& -\left(r_q-
\frac{\phi}{\phi_s}(r_q-r_se^{{\phi_s}\over \ze})\right){e^{{-\phi\over
\ze}}\over \ze}
-\frac{1}{\phi_s}(r_q-r_se^{{\phi_s}\over \ze})e^{{-\phi}\over \ze}\nonumber \\
&=& -{r\over \ze} -\frac{r_q-r_se^{{\phi_s}\over
\ze}}{\phi_s}e^{{-\phi\over \ze }}
\eea
so that, along $\gamma$, 
\be
g(\dot{\gamma},\dot{\gamma})  = 4 r^2(-\ze^2\dot{r}^2
+r^2\dot{\phi}^2)=4r^2\dot{\phi}^2 \ze^2 \left( - (\frac{dr}{d\phi})^2
+{r^2\over \ze^2}\right)
\ee
Since $r_q-r_se^{{\phi_s}\over \ze}>0$ we have
$\left|\frac{dr}{d\phi}\right|>{|r|\over Z}$, so that the tangent to $\gamma$ is
everywhere timelike. We can choose the direction of parameterisation so that 
$\gamma$ is future directed. Thus
all points in $L_q$ belong to $\pa$. Moreover, no point outside $L_q$
belongs to $\pa$, since the curve would have to cross the small heart
boundary at some point $s$; but according to the local lightcone at $s$ any
vector in its tangent space $T_{s}$ pointing into $L_q$ is either spacelike or
past directed.

The causal past and future of $q$ in this case are simply the closure of their
chronological analogues since  the bounding curves of the lightcones are 
always geodesics through $q$. 

That the above geometry is causally continuous follows from the
general result lemma \ref{yarmulke.lemma}. But the reader can
verify it graphically by finding the intersections of the past of all points to the
future of $q$ and vice-versa.

\subsection{The ``chronos'' of the Morse point}
\label{chronos.sec}

Violation of causal continuity in the trousers occurs at points on the null
geodesics that approach the origin.  Looking back at figure 
(\ref{tpart.fig})
one is inclined to regard these lines as the null cone of the critical
point $p$, with $\mathcal{F}$ constituting its chronological future and
$\mathcal{P}$ its chronological past. Although this is indeed the case
in some scheme where the critical point is retained, we have excised it for
present purposes and therefore need a few more definitions to give
$\mathcal{F}$ and $\mathcal{P}$ a more formal status.

An {\it indecomposable past set (IP)\/} in a geometry $(M,g)$ is a subset $W$
of $M$ possessing the following two properties: $1.\; I^-(W) = W$, a set
with this property is deemed to be a {\it past set}.  $2.\; W$ cannot be
decomposed as the union of two proper past sets.  An indecomposable future
set (IF) is defined dually.

The chronological past and future of any point in a Lorentzian spacetime
are the standard examples; that is for any $q$, $I^-(q)$ is an IP and
$I^+(q)$ is an IF. The IP's and IF's that are not of this form were
introduced to probe the causal boundary of spacetime. They are called
terminal indecomposable pasts (TIP) and futures (TIF)\cite{hawking73}. 

It can be shown that a set $W$ is a TIP if and only if there is a future
inextendible \footnote{A causal curve in some spacetime is said to be future
(past) inextendible if it has no future (past) endpoint. Inextendibility of
a geodesic shows up as it approaches the edge of spacetime or a singularity.}
timelike curve $\gamma$ such that $W=I^-(\gamma)$. In general, a whole
class of such curves generates the same TIP
\footnote{Hawking and Ellis identify
IP's with IF's so that  the boundaries of space, both at infinity or
singularities, are in some sense attached to the spacetime. They take the
classes of such past and  future sets to be the elements among which a new
causal space is defined. It may be interesting to pursue this 
approach in our cases and
confirm that it  leads to the same conclusions about causal continuity.}. 
These concepts have an immediate application to our elementary
cobordisms, since the removed critical point is, in this sense, a boundary of
the spacetime.  

An illustration is furnished by the $(1+1)$ dimensional
examples from the previous sections.  In the trousers it is 
straightforward to show there are two
distinct TIP's, $\mathcal{P}_1$ and $\mathcal{P}_2 $, corresponding to each
of the future inextendible timelike curves
$\omega_{1}(t)=(x(t),y(t))=({\epsilon\over 2}e^{-t},0)$ and
$\omega_{2}(t)=(-{\epsilon\over 2}e^{-t},0)$, $t\in[0,\infty)$. 
Similarly, there are two distinct TIF's,
$\mathcal{F}_1$ and $\mathcal{F}_2$, generated by the past inextendible
timelike curves $\gamma_{1}(t)=(0,{\epsilon\over 2} e^t)$ 
and $\gamma_{2}(t)=(0, -{\epsilon\over 2}e^t)$, $t\in(-\infty,0]$.

In the 1+1 yarmulke, on the other hand, every radial geodesic is a past
inextendible timelike curve. In fact they are all equivalent as future set
generators, since any such curve $\gamma$ generates $I^+(\gamma) = \N$, the
whole punctured disc.  Thus we say that the TIF representing the singularity
is the entire $\N$. But no timelike curve is future inextendible towards the
origin: the TIP associated with the creation point is empty. 
So in this case ${\mathcal F}= \N$ and ${\mathcal P}=\emptyset$, as expected.

To summarise, we see that in these $2$-dimensional geometries we can
 use inextendible timelike curves whose endpoint would be the origin
 to view the singularity as an ideal point \cite{geroch72} in the
 spacetime and thus talk about its past $\mathcal{P}$ and its future
 $\mathcal{F}$. The topology of these sets plays a role in determining
 the causal structure around the critical point.

 In the causally discontinuous $(1+1)$ trousers both $\mathcal{P}$ and
$\mathcal{F}$ consist of two disconnected components.  Moreover
$\mathcal{F}$ ``separates'' $\mathcal{P}$, in the sense that any curve
joining the two disconnected components of $\mathcal{P}$ necessarily
traverses $\mathcal{F}$ . Similarly $\mathcal{P}$ separates
$\mathcal{F}$. The discontinuity of $I^-$ is manifest in that the past
of any point on $\partial { \mathcal{F}}$ intersects only one of the
two components of $\mathcal{P}$, while any point in $\mathcal{F}$
contains the whole of $\mathcal{P}$ in its past. Dual statements can
be made regarding the discontinuity of $I^+$.

In the causally continuous $(1+1)$ yarmulke and its time reverse,
both $\mathcal F$ and $\mathcal P$, respectively, are trivially connected.

This seems to indicate that the source of causal discontinuity for the
neighbourhood geometries resides on the boundaries of the TIFs and
TIPs, and suggests a scheme to investigate the neighbourhood of
a critical point of arbitrary index $\lambda$ in general dimension
$n$.

\section{Causal continuity in  higher dimensional Morse geometries}

We now examine the causal continuity of the neighbourhood geometries with
arbitrary $\lambda$ for $n\ge 3$.   
For the yarmulke neighbourhoods 
lemma \ref{yarmulke.lemma} shows that they are 
causally continuous without further work. Thus in the remainder
of this section we impose the condition $\lambda \ne 0,n$. 
The causal structure in these 
cases can be analysed by  identifying 
higher dimensional analogues of the TIP's and TIF's
discussed in  Section \ref{chronos.sec}
and appropriately combining the causal
structures of the  $2$-dimensional spacetimes studied in Section
\ref{2d.sec}.

\subsection{TIP's and TIF's in higher dimensions.}\label{generaltiptif.section}

Guided by the form of $\mathcal{F}$ and $\mathcal{P}$
in the trousers, we start by guessing at their analogues in the higher
dimensional neighbourhood geometries. For the moment, we mark these sets
with  a tilde, but then show that they are in fact TIP's and TIF's 
associated
with the singularity. We define them in the spherical coordinates used in equation
 (\ref{sphmetric.eq}). The 
angular coordinates (including the discrete 
angle if $\lambda =1$ or $ n-1$) 
are understood to vary over all their possible values.
\bea
\tilde{\mathcal{P}} &\equiv& \{ q\in \N : r_q < m_1\rho_q\} \nonumber \\
\nonumber \\
\tilde{\mathcal{F}} &\equiv& \{ q\in \N : r_q> m_2\rho_q\}
\label{tildePF.eq}
\eea
where $\rho_q$ is the $\rho$ coordinate of $q$, {\it etc.}, and $m_1$ and
$m_2$ are as in (\ref{m1m2.eq}). 

Let us define the quadrant $\chi_q$ to be the set of points with the same
angular coordinates as $q$. The warped product form of
(\ref{sphmetric.eq}) with respect to the pair $(\rho, r)$, ensures
that the geodesics in this quadrant are geodesics of the full
metric. 

\begin{figure}[ht]
\centering
{\resizebox{!}{3in}{\includegraphics{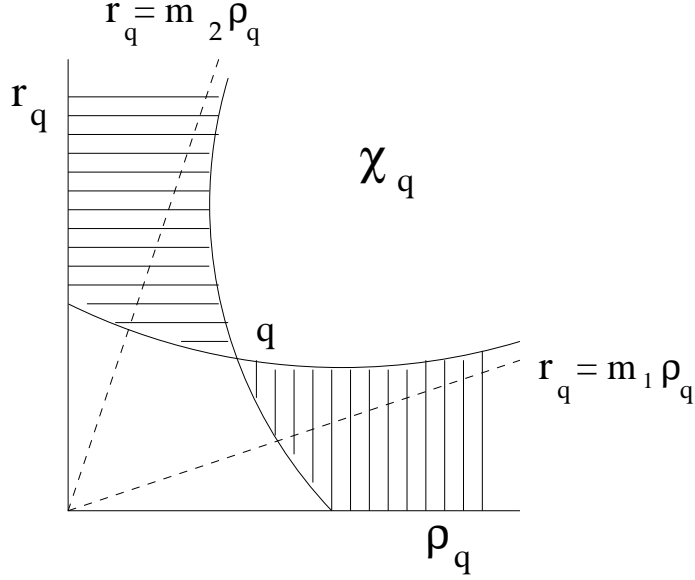}}}
\caption{{\small
$\chi_q$. $q\in S$  and
$I^+(q, \chi_q)$ ($I^-(q, \chi_q)$) is the horizontally
(vertically) shaded region. 
}\label{higher.fig}}
\end{figure}

\begin{figure}[ht]
\centering
{\resizebox{!}{3in}{\includegraphics{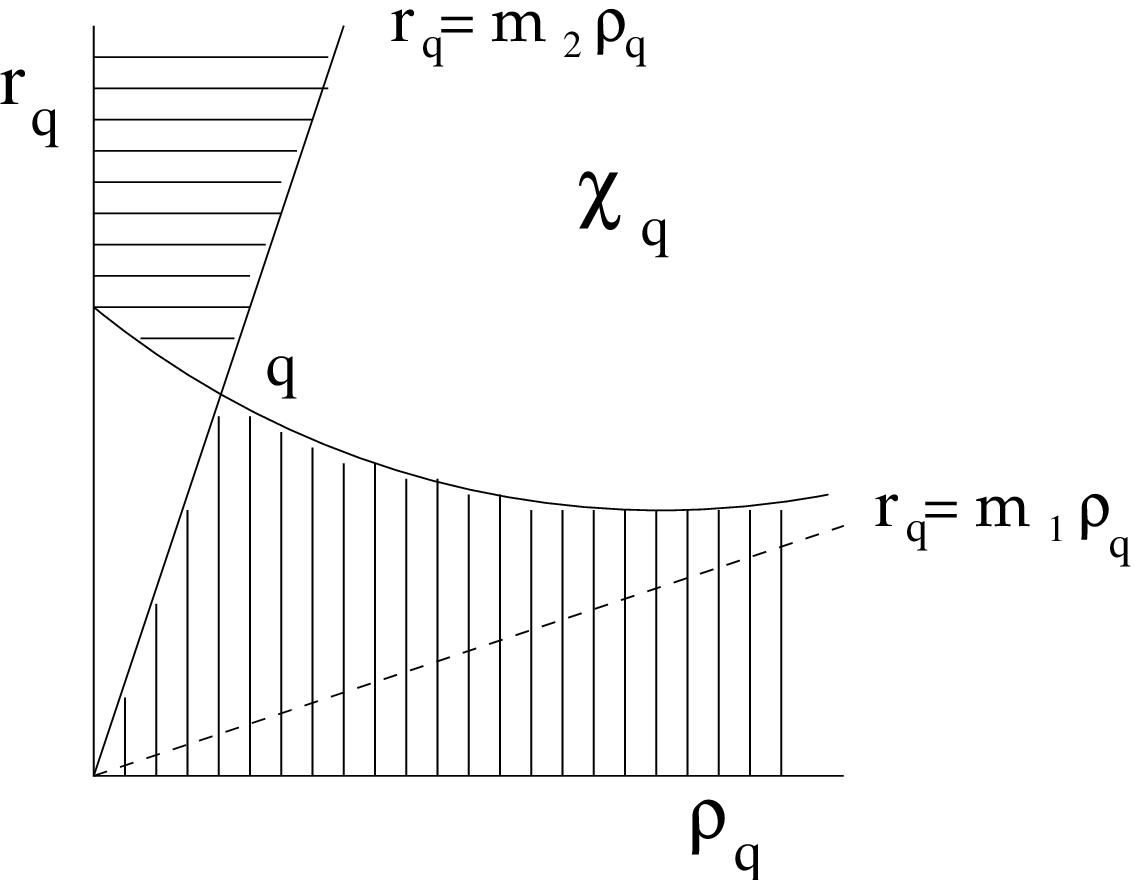}}}
\caption{{\small $\chi_q$. $q \in \partial\FU$.
$I^+(q, \chi_q)$ ($I^-(q, \chi_q)$) is the horizontally 
(vertically) shaded region.
}\label{tif.fig}}
\end{figure}

\begin{figure}[ht]
\centering
{\resizebox{!}{3in}{\includegraphics{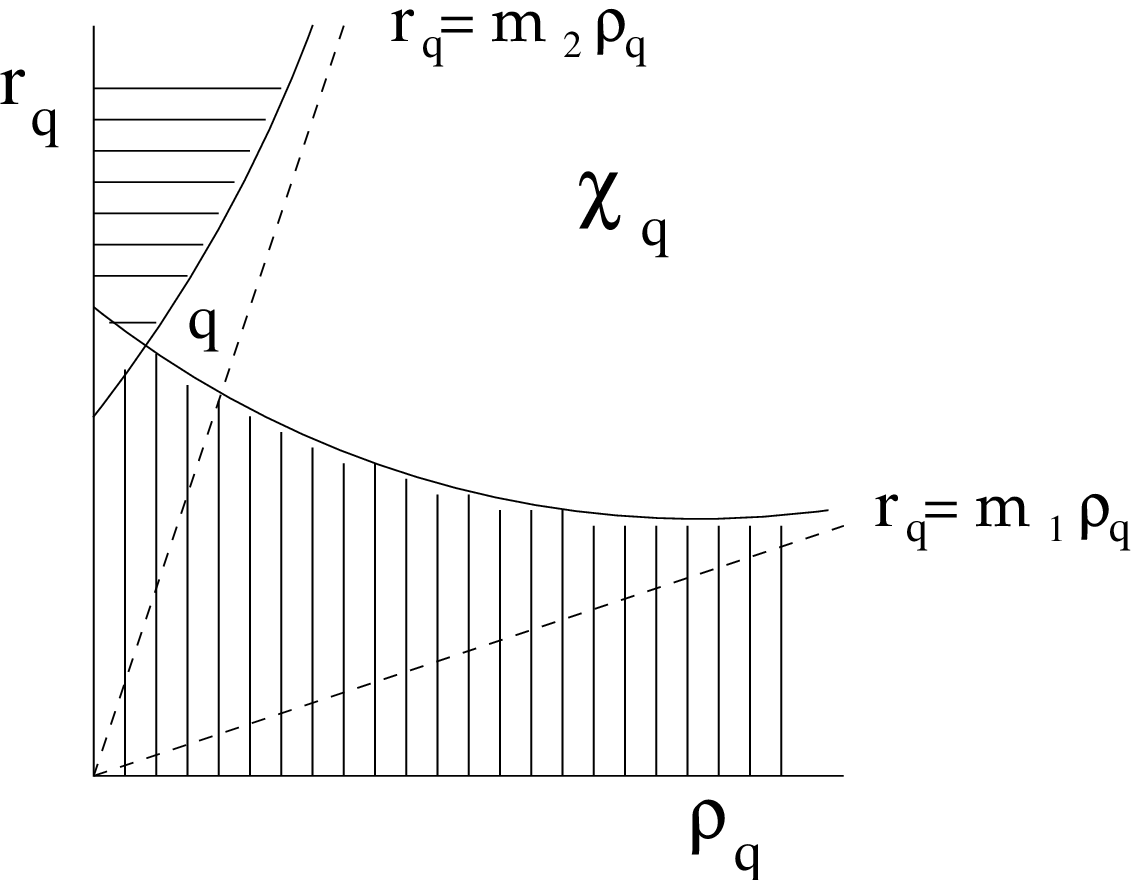}}}
\caption{{\small $\chi_q$. $q\in \FU$ and  
$I^+(q, \chi_q)$ ($I^-(q, \chi_q)$) is the horizontally 
(vertically) shaded region.
}\label{chi.fig}}
\end{figure}

We introduce a projection operator $P_q$ which acts on points and curves by
projecting them into the quadrant $\chi_q$, so that for $x \in \N$ with $x =
(\rho_x, \Theta_x, r_x, \Phi_x) $, $P_q x= (\rho_x,
\Theta_q, r_x, \Phi_q)$. 
Similarly, if $\gamma(t)$ is a curve with coordinates $\gamma(t) =
(\rho(t), \Theta(t), r(t), \Phi(t))$ then $P_q \gamma(t) = (\rho(t),
\Theta_q, r(t), \Phi_q)$.  A timelike curve $\gamma$ projects to a
timelike curve $P_q \gamma$ in the quadrant and a causal curve
projects to a causal curve. It follows that
$I^{\pm}(q,\chi_q)=I^{\pm}(q)\cap\chi_q$ and moreover that for any
point $y$ in $I^{\pm}(q)$ its projection $P_q y$ to $\chi_q$ must lie in
$I^{\pm}-(q,\chi_q)$.  Hence, the causal structure in $\chi_q$ is that of
the upper right quadrant of the trousers and this is illustrated in
figures \ref{higher.fig}--\ref{chi.fig} in which we give the
chronological pasts and futures of three representative points.

 We further define the set $\tilde{\mathcal{P}}_q$ to be the intersection
$\tilde{\mathcal{P}}\cap\chi_q$ and $\tilde{\FU}_q=\tilde{\FU} \cap \chi_q$.

If $\lambda$ is neither $1$ nor $n-1$, 
$\tilde{\mathcal P}$ and $\tilde{\mathcal F}$ are connected sets. 
However when $\lambda =1$ 
$\tilde{\mathcal{P}}$ comprises two disjoint components
$\tilde{\mathcal{P}}_1$  which has $\theta_0 = 0$ 
and $\tilde{\mathcal{P}}_2$  with $\theta_0=\pi$.
Moreover $\tilde{\mathcal{F}}$, which is connected ($n\ge 3$), then separates
$\tilde{\mathcal{P}}$ in the sense that any curve between
$\tilde{\mathcal{P}}_1$ and $\tilde{\mathcal{P}}_2$ necessarily intersects
the $r=0$ hyperplane, which is contained in
$\tilde{\mathcal{F}}$. When $\lambda=n-1$, in the time-reversed geometry,
$\tilde{\mathcal{P}}$ separates $\tilde{\mathcal{F}} = \tilde\FU_1
\coprod \tilde\FU_2$ similarly.

Now we define a terminal past set and a terminal future set, generated by
inextendible timelike curves that approach the Morse point. Let the curve 
$\omega_a$ in $\N$ be given in polar coordinates by
\be
\omega_a(t) =  ({\epsilon\over 2} e^{-t}, 
\Theta_a, 0, 0), \qquad  t \in [0,\infty)\ ,
\ee
where $\Theta_a$ denotes some
fixed point on the $(\lambda-1)$ sphere and $\Phi$ is strictly undefined
 since $r=0$  though we have written it as $0$, and the curve $\gamma_b$ be 
given by
\be
\gamma_b(t) = (0,0, {\epsilon \over 2}e^t, \Phi_b), \qquad 
t \in (-\infty, 0]\ .
\ee 
These are both timelike curves; $\omega_a$ is future inextendible and
$\gamma_b$ is past inextendible. 

For $\lambda \ne 1$, we define $ \PA \equiv I^-(\omega_a)$. We show 
that $\PA$ does not depend on the choice of $\Theta_a$. Consider 
$\omega_{a'}(t) = ({\epsilon\over 2}e^{-t}, \Theta_{a'}, 0,0)$ and let 
$\PA' \equiv  I^-(\omega_{a'})$.  Consider a point $z$ in $\PA$ 
so that there exists a future directed timelike curve from $z = (\rho_z,
\Theta_z,r_z,\Phi_z)$ to some point of $\omega_a$. The lines 
$\omega_a$ and $\omega_{a'}$ lie in a  unique 2-plane on which the 
induced metric is that of the  $(2,0)$ yarmulke and in 
which the two curves are radial timelike geodesics. It follows from 
the analysis of Section  \ref{yarmulke.subsec} 
that within that 2-plane there are future directed timelike
spiraling curves that start at any given point of $\omega_a$ and 
end at some point of $\omega_{a'}$. Therefore there's a future
directed timelike curve from $z$ to some point of $\omega_{a'}$ and 
so $z\in \PA'$. Reversing the roles of $a$ and $a'$ we also have
 $\PA' \subset \PA$ and so $\PA' = \PA$. 

Similarly for $\lambda \ne n-1$ we define $\FU \equiv 
I^+(\gamma_b)$. An identical argument shows that $\FU$ does not 
depend on the choice of $\Phi_b$: any $\gamma_b$ generates it. 

When $\lambda = 1$, instead of one generator for $\PA$ we require two,
one for each possible value of the $S^0$ coordinate and assign them fixed labels
$1$ and $2$:
\bea
\omega_1(t) &=& ({\epsilon\over 2} 
e^{-t}, \theta_0 = 0, 0, 0), \qquad t \in [0,\infty)\\
\omega_2(t) &=& ({\epsilon\over 2}e^{-t}, 
\theta_0 = \pi, 0,0), \qquad t \in [0,\infty)\ .
\eea
Then let $\PA_i \equiv
 I^-(\omega_i)$, $i = 1,2$ and $\PA \equiv \PA_1 \coprod \PA_2$. 

Finally when $\lambda = n-1$ there are two generators for $\FU$:
\bea
\gamma_1(t) &=& (0,0, {\epsilon\over2}e^t, \phi_0=0) \qquad t \in(-\infty,0]\\
\gamma_2(t) &=& (0,0, {\epsilon\over2}e^t, \phi_0=\pi)\qquad t \in(-\infty,0]
\eea
with $\FU_i \equiv I^+(\gamma_i)$, $i = 1,2$ and $\FU 
\equiv \FU_1\coprod \FU_2$.

We now show that $\PA = \tilde \PA$. The proof that $\FU = \tilde \FU$ is 
analogous. There are two cases. 

a) $\lambda \ne 1$ 

Let $z\in \PA$. Take the 
curve $\omega_z$ with $S^\lambda$ coordinate $\Theta_z$ as the 
generator of $\PA$. Consider the 2-d quadrant $\chi_z$. 
$\chi_z$ has the causal structure of the 1+1 trousers as 
represented by figures \ref{higher.fig}--\ref{chi.fig}.
In $\chi_z$, $\omega_z$ is the $\rho$-axis (horizontal or past axis).  $z\in 
I^-(\omega_z, \chi_z)$ and so it must 
satisfy $r_z < m_1 \rho_z$ and so $z \in \tilde \PA$.
Now suppose $u \in \tilde \PA$. Let $\omega_u$ be a generator
of $\PA$ with $\Theta = \Theta_u$. Then $u \in I^+(\omega_u, \chi_u)$
by the causal structure of $\chi_u$ and so $u \in \PA$.

b) $\lambda =1$

Wlog let $z \in \PA_1$. Then
$\omega_1$, the generator of $\PA_1$ is the $\rho$-axis in the 
quadrant $\chi_z$ and 
$r_z <m_1 \rho_z$ by the same argument as above and so 
$\PA_1 \subset \tilde \PA_1$. Finally let $u \in \tilde \PA_1$. 
In $\chi_u$, $u \in I^-(\omega_1, \chi_u)$ and we are done.  

We again define the sets 
\bea
\partial\mathcal{P} &\equiv& \{x\in \N : r_x = m_1 \rho_x \} \nonumber\\
\partial\mathcal{F} & \equiv& \{x\in \N: r_x = m_2 \rho_x \} \nonumber\\
\mathcal{S} &\equiv& \N - (\mathcal{P}\cup\mathcal{F}\cup \partial\mathcal{P} 
\cup \partial\mathcal{F} ) 
\eea
and let $\mathcal{F}_q \equiv \mathcal{F} \cap \chi_q$, {\it etc.}

\begin{claim}
\label{PandF.claim} 
(i) If $q \in \mathcal{F}$, then 
$\mathcal{P} \subset I^-(q)$. (ii) If $q\in\partial \mathcal{F}$ then $I^+(q)\subset 
\mathcal{F}$. (iii) If $q\in\partial \mathcal{F}$, then 
(a) $\;\lambda\neq 1$ implies 
$\mathcal{P}\subset \pa$, while (b) $\lambda =1$ implies either 
$I^-(q)\cap\mathcal{P}_2=\emptyset$ or $I^-(q)\cap \mathcal{P}_1=\emptyset$.
\end{claim}

\noindent
{\bf Proof:} (i) Let 
$q\in \FU$ and 
$z \in \PA $.
$\FU$ is generated by $\gamma_q$ with $\Phi=\Phi_q$ and
$q\in I^+(\gamma_q)$. $\PA$ is generated by 
$\omega_z$ with $\Theta=\Theta_z$ and
$z\in I^-(\omega_z)$. The curves $\omega_z$ and $\gamma_q$ lie in the 
2-d quadrant defined by $\Theta=\Theta_z$ and $\Phi = \Phi_q$  in which 
$\omega_z$ is the $\rho$-axis and $\gamma_q$ 
is the 
$r$-axis. By the causal structure of  $\chi_q$, for every point $q'\in \gamma_q$
we have $\omega_z \subset I^-(q')$. Hence result.

(ii) Let $q \in \partial\FU$ and $z \in I^+(q)$. 
Consider the projection $P_q z$ into $\chi_q$ which must lie
in $I^+(q,\chi_q)$. This gives us $r_z > m_2 \rho_z$ and so 
$z \in \FU$. Note that we didn't actually need to explicitly prove this point, since in general, for any future set $A=I^+(B)$ we have $I^+(\partial A)\subset A$ \cite{penrose72}.

(iii) 
Let $q  \in \partial\FU$.

\noindent(a)
$\lambda \neq 1$.  Let $z \in \PA$. $\PA$ is generated by $\omega_q$ 
with $\Theta=\Theta_q$ and $z \in I^-(\omega_q)$.
Now $\omega_q$ is the 
$\rho$ axis of the quadrant $\chi_q$
and again the causal structure in this quadrant implies that 
$\omega_q \subset I^-(q)$ and so $z \in I^-(q)$.

\noindent (b) $\lambda = 1$. Suppose wlog $\theta_0(q) = 0$ and let 
$z \in I^-(q)$. If $\theta_0(z) = \pi$ then there exists a future
directed timelike curve from $z$ to $q$ which passes through
$\rho = 0$. Considering the projection of this curve into 
$\chi_q$ shows this to be a contradiction.
So $\theta_0(z) = 0$ and $I^-(q) \cap \PA_2 = \emptyset$.
$\Box$ 

\subsection{Causal continuity in neighbourhood geometries of general index and
dimension}

We use the following notation to denote scale transformations by a real
number $a$: given a point $q=(\rho_q,\Theta_q,r_q,\Phi_q)$ and a curve
$\gamma(t)=(\rho(t),\Theta(t),r(t),\Phi(t))$, we write $aq$ for the point
$aq=(a\rho_q,\Theta_q,ar_q,\Phi_q)$ and $a\gamma$ for the curve
$a\gamma(t)=(a\rho(t),\Theta(t),ar(t),\Phi(t))$. Notice that the timelike
character of a curve is preserved under this scaling. We refer to
$R^2=\rho^2 + r^2$ as the squared distance from the origin.

\begin{lemma}\label{PFcc.lemma} For any index $\lambda $, causal continuity
holds at all points $q\in\mathcal{P}\cup\mathcal{F}$. 
\end{lemma}

\noindent
{\bf Proof:} 
Let $q \in \FU$. 
First, we prove that $\cpa=\pa$. Suppose not. 
Then $\cpa \not\subset \pa$ and  
by claim \ref{notjustclosure.claim} (appendix) there is a point $y$ and a 
neighbourhood $U_y$ of $y$ such that  $U_y \subset \cpa$ and 
$U_y \cap I^-(q)= \emptyset$. 
Define the sequence of points $q_k = a_k q$ 
where $a_k = (1 + {\delta\over k})$, $k = 1,2,\dots$
 and $\delta >0$ is small enough that $q_1 \in \N$.  The $q_k$ 
tend to $q$ and lie along the radial timelike line from the origin through 
$q$. So $q_k \in I^+(q)$, $\forall k$. Thus there exists a future directed 
timelike curve $\gamma_k$ from $y$ to $q_k$. Let 
$\gamma'_k = a_k^{-1}\gamma_k$,  again a future directed 
timelike curve. The final point of each of these scaled curves is $q$ 
and the initial point of $\gamma_k'$ is $y_k = a_k^{-1} y$.  Choose
$k$ large enough so that $y_k \in U_y$. This is a contradiction. 

To prove that $\cfu=\fu$ we again assume it does not hold and so
there is a point $y$ and a 
neighbourhood $U_y$ of $y$ such that  $U_y \subset \cfu$ and 
$U_y \cap I^+(q) = \emptyset$. Let $q_k = b_k q$ where $b_k = 1-{\delta\over k}$,
$k = 1,2,\dots$. Then each $q_k \in I^-(q)$ and so $\exists$ a future 
directed timelike curve $\gamma_k$ from $q_k$ to $y$. Let $\gamma_k'
= b_k^{-1} \gamma_k$ which is again future directed and timelike. The 
initial point of each $\gamma_k'$ is $q$ and the final point of
$\gamma_k'$ is $y_k = b_k^{-1} y$. 

We must now check that for large enough $k$, the curve $\gamma_k'$ lies 
entirely within the disc $\N$. $\exists$ $K>0$ such that $k>K$ implies that 
$y_k \in \N$. Also we have $q_k \in \FU$, $\forall k$. By claim
(\ref{Rincreases.claim}) (appendix)
we therefore know that $R^2$ reaches its maximum along 
$\gamma_k'$ at its future endpoint $y_k$ and so $\gamma_k'$ remains in 
$\N$ for $k>K$. Now choose some $k> K$ large enough so that $y_k \in U_y$
which is a contradiction. Hence the result.

Causal continuity at points in $\mathcal{P}$ is proved similarly. $\Box$

\begin{lemma}\label{Scc.lemma} For any index $\lambda$, causal continuity
holds at any point $q\in\mathcal{S}$. 
\end{lemma}

\noindent
{\bf Proof:} We show that $\pa = \cpa$. Suppose not.
Then, as before,
 there exists a $y \in \cpa$ with a neighbourhood $U_y$ contained in $\cpa$ 
which doesn't intersect $\pa$.  Choose some 
sequence of points $q_k \in I^+(q,\chi_q)\cap S$, $k = 1,2,\dots$,
that converges to $q$ from the 
future so that $q_k \in I^+(q_{k+1})$. For example
they could lie on the timelike hyperbola $r\rho = r_q\rho_q$ 
through $q$ in $\chi_q$.
There exist future directed timelike curves, $\gamma_k$ from $y$ to $q_k$ 
$\forall k$. We plan to use the CCLT to construct a causal curve
from $y$ to $q$ and to do so we must identify a compact region in which all 
(or at least infinitely many of) the $\gamma_k$ are contained. The set of all
the $\gamma_k$ is bounded away from the origin since 
each $\gamma_k$ must lie in $I^-(q_1)$ which is seen to be bounded away from the 
origin by considering $I^-(q_1) \cap \chi_{q_1}$. 

However it might be that the 
set is not bounded away from the edge of the punctured disc, $N_\epsilon$.
So consider the geometry with metric (\ref{sphmetric.eq})
on the larger punctured disc $N_{2\epsilon} \equiv D_{2\epsilon} - \{p\}$. 
Our neighbourhood
geometry $(\N, g)$ is embedded in it in the obvious way. 
In $N_{2\epsilon}$ we can find a compact set in 
which all the $\gamma_k$ lie and which is $C = \overline{{D}_\epsilon }
- D_{\epsilon'}$ where $\overline{{D}_\epsilon}$ is the closed (in $\mathR^n$) 
ball of radius
$\epsilon$ and $\epsilon'>0$ is small enough.

The CCLT now implies that $\exists$ a causal curve in $C$ 
from $y$ to $q$. 

This is true for all points $y' \in U_y$. So that $U_y \subset J^-(q, C)$. 
Since $Int(J^-(q,C)) \subset Int(I^-(q,C)) = I^-(q,C)$ this implies that 
there exists a timelike curve, $\gamma$, from $y$ to $q$ in $C$. If the radial 
distance $R$ is bounded away from $\epsilon$ along $\gamma$  we are done, 
since then $\gamma$ lies in $\N$. 
So suppose it is not. We construct a new curve $\tilde\gamma$ by rescaling 
$\gamma$ down away from the boundary almost everywhere, except in a small 
neighbourhood of $q$ where we fix it so that it still ends at
$q$. 

More precisely let $U_q \subset \N$ be a neighbourhood of $q$ and let 
$s \subset U_q$ be a point on $\gamma$. In particular $s \in I^-(q)$. 
Choose $\delta >0$ small enough so that $s' = (1-\delta)s$ is also 
in $I^-(q)$ and $y' = (1-\delta)y$ is in $U_y$. Then the initial 
segment of $\tilde\gamma$ is $(1-\delta)\gamma$ from $y'$ to 
$s'$ and then we smooth it into a timelike curve from $s'$ to $q$. 
$\tilde\gamma$ then lies entirely within $\N$ and so $y' \in I^-(q)$ 
which is a contradiction. 

$\fu = \cfu$ is proved similarly. $\Box$

The only potential obstructions to global causal continuity 
therefore lie in the 
remaining
region $\partial\mathcal{F}\cup\partial\mathcal{P}$. 
Combining the method of the previous lemma and the results gathered in
the last section we can prove.
\begin{lemma}\label{partialPFcc.lemma}
(a) When $\lambda \ne 1$ causal continuity holds at all points 
$q\in \partial\FU$. (b) When $\lambda = 1$, if $q\in \partial\FU$ then 
$\pa \ne \cpa$. (c) When $\lambda \ne n-1$ causal continuity holds at all
points $q\in \partial\PA$. (d) When $\lambda = n-1$, if $q\in\partial\PA$ then
$\fu\ne\cfu$. 
\end{lemma}

\noindent
{\bf Proof:} Consider
$q\in \partial\mathcal{F}$. 

\noindent 
(a) $\lambda \ne 1$. We first show that $\cpa = \pa$. Suppose
not. Then as usual, there is a point $y$ with a neighbourhood, $U_y$,
such that $U_y \subset \cpa$ and $U_y\cap \pa = \emptyset$. First $y\not\in
\PA$ by claim \ref{PandF.claim} (iiia); also $y\not\in\partial\PA$ since
otherwise some other point in $U_y$ would be in $\PA$. Secondly  $y\not\in
\FU$, since otherwise considering a sequence of points $q_k\rightarrow
q$, with $q_k\in I^+(q,\chi_q)$ and the fact that $P_q y \in I^-(q_k,
\chi_q)$ $\forall k$ would lead to a contradiction with the known
causal structure of $\chi_q$; again $y\not\in\partial\FU$ since
otherwise $U_y\cap\FU \ne \emptyset$ and the same contradiction would
arise.  Finally if $y$ lies in $S$ then using arguments similar to
those of the previous lemma we would obtain a contradiction also.
 
\noindent (b) $\lambda=1$. Wlog let $\theta_0(q) = 0$. Let $y\in \PA_2$. 
Then $x\in I^+(q) \Rightarrow x \in \FU$ by claim \ref{PandF.claim}(ii)
$\Rightarrow y\in I^-(x)$ by claim \ref{PandF.claim}(i) and similarly for any
point $y'$ in a neighbourhood, $U_y$, of $y$ such that $U_y \subset \PA_2$.
Thus $y\in \cpa$. But $y\not\in I^-(q)$ by the proof of claim 
\ref{PandF.claim}(iiib).

Parts (c) and (d) are proved similarly. $\Box$

Putting together the partial results in lemmas \ref{yarmulke.lemma},
\ref{tcd.lemma} \ref{PFcc.lemma}, \ref{Scc.lemma} and
\ref{partialPFcc.lemma}, we have a proof of the following proposition. 

\begin{proposition}\label{ccinDnbhood.proposition}
Let $(\N,g)$ be the neighbourhood geometry type $(\lambda, n-\lambda)$.
 Then

(i) If $\lambda \neq 1, n-1$,  $(\N,g)$ is causally continuous.

(ii) If $\lambda =1$ or $\lambda =n-1$, $(\N,g)$ is not causally
continuous.
\end{proposition}

\section{Discussion}

We have made progress on the way to proving the Borde-Sorkin
conjecture that causal continuity in Morse geometries for topology
change is associated only with indices $\lambda \ne 1, n-1$.  In
particular we have proved that the conjecture holds for certain
geometries that are neighbourhoods of single Morse points. We have
also proved some more general results. The causal continuity of the
yarmulke spacetimes has been demonstrated for {\it any} Morse metric
on these cobordisms. We have also shown that any strongly causal
compact product spacetime must be a Morse metric. This shows that the
class of Morse spacetimes is in fact quite general and hence supports
the proposal to sum over Morse metrics in the SOH.

In order to obtain a full proof, first our main results in the
neighbourhood need to be extended to more general metrics than those built from
the Cartesian flat Riemannian auxiliary metric.  The next step would be to
understand how the causal properties of the individual neighbourhoods
affect the causal properties of the entire spacetime. We address these
issues in  a forthcoming paper \cite{cdproof}.  

We mentioned in the introduction our intention, eventually, to consider the
Morse points as part of the spacetime and not to excise them. A causal
order that includes the critical points is desirable, especially since it
does not seem plausible that isolated points should be relevant in the
quantum context. Most simply we can add in the degenerate point, $p$, and
extend the causal relation by hand as follows: the causal past and future
of $p$ are taken to be the closure of the union of respectively the TIPs
and TIFs associated with $p$. For our neighbourhood spacetimes, then
$J^-(p)$ would be $\overline\PA$, and $J^+(p)=\overline\FU$, where the
closure is taken in the unpunctured disc. The full causal relation on the
spacetime is then completed by transitivity.

Though simple, this prescription may appear a little ad hoc. We
believe, however, that the causal order obtained in this way coincides
with a robust generalisation of the usual causal relation proposed in
\cite{erik95}.  This new relation, called $K$, is defined in terms
of the chronological relation $I$, which can be extended to the Morse
geometries by using the strict definition for chronology, i.e., by
putting $I^\pm(p)=\phi$. One can immediately verify that the 
causal relation proposed above is the same as $K$ for the
neighbourhood geometries. Our trivial extension of the chronology to
the critical points seems justified by the robustness of $K$, which is
indifferent to the presence of isolated points in a spacetime. A
remarkable feature of this new setting is that the property of causal
continuity appears as the condition that the pair $(I,K)$ constitute a
genuine causal structure, in the axiomatic sense of \cite
{kronheimer68}. Details of this analysis will appear elsewhere
\cite{kpaper}.

Even if further work is required before we can understand the physical
relevance of our results, in particular the effect of causal
(dis)continuity on the propagation of quantum fields, we can already
make an interesting observation: if the universe did begin in a big
bang that could be described in its earliest moments by a Morse metric
with an index 0 point, then the causal structure of the yarmulke is
such that there are no particle horizons: all points on a given level
surface of the Morse function have past points in common. That would
mean that there would no longer be a cosmological horizon problem.

\section{Acknowledgments}
We would like to thank Andrew Chamblin and Bernard Kay for interesting
discussions.  AB thanks the Inter-University Center for Astronomy and
Astrophysics, Pune, India, and the Institute of Cosmology at Tufts
University for their hospitality, and the Faculty Research Awards Committee
at Southampton College for partial financial support. SS was supported by a
postdoctoral fellowship from T.I.F.R., Mumbai. HFD was supported by an
EPSRC Advanced Fellowship. RSG is supported by a Beit Fellowship. RDS
was partly supported by NSF grant PHY-9600620 and by a grant from
the Office of Research and Computing of Syracuse University.

\section{Appendix}
We prove two results that we use in the text.
\begin{claim}
 \label{notjustclosure.claim} If $\exists y \in \downarrow I^+(x)$
such that $y \notin I^-(x)$ then there is also exists a $z \in \downarrow
I^+(x)$  such that  $z \notin \overline{I^-(x)}$. Dually,  if $\;\uparrow
I^-(x) - I^+(x)$ is not empty, then there is a point $z$ in $\uparrow
I^-(x)$ with    $z\notin \overline{I^+(x)}$.
\end{claim}

\noindent
{\bf Proof:} Suppose not, namely suppose
$\downarrow I^+(x)-I^-(x)$ is not empty and all of its points lie in
$\partial I^-(x)$. Pick one such $y$.
By the obvious 
generalisation of Proposition 6.3.1\cite{hawking73} to a Lorentzian metric in 
general dimension, the boundary $\partial I^-(x)$ is an $n-1$ dimensional 
submanifold, so that any ball centered on the boundary
intersects both $I^-(x)$ and the complement of $\overline{I^-(x)}$.
Thus every neighbourhood of $y$ intersects
the complement of $\overline{I^-(x)}$; hence no neighbourhood of $y$ is 
contained in $\downarrow I^+(x)$,  a contradiction. $\Box$

\begin{claim}\label{Rincreases.claim} Any future directed timelike
curve, $\gamma$ in $\N$ that begins in $\FU$ remains in $\FU$
and moreover the function $R^2=\rho^2 + r^2$ 
increases monotonically along it.
\end{claim}

\noindent
{\bf Proof:} 
Let $\gamma$ have initial point $q\in \FU$. Consider the projected 
curve $P_q \gamma$. It has the same $r$ and $\rho$ behaviour as
$\gamma$ and so if we prove the result for $P_q \gamma$
then it follows for $\gamma$ itself.

$P_q\gamma$ lies in $\chi_q$ and starts at $q \in\FU_q$. The 1+1 
trousers causal structure of $\chi_q$ shows that $P_q\gamma$ 
remains in $\FU_q$. 
So we have 
\begin{enumerate}
\vspace{-4mm}
\item $r\dot{r}-\rho\dot{\rho}>0$
\item $(r\dot{r}-\rho\dot{\rho})^2>(r\dot{\rho}+\rho\dot{r})^2$.
\item $r>m_2\rho$
\vspace{-4mm}
\end{enumerate}
 Their combination forces $r\dot{r}+\rho\dot{\rho}$ to be positive 
along $P_q \gamma$. We check this for the two possible signs of 
$r\dot{\rho}+\rho\dot{r}$  in condition $2$.
a) If $r\dot{\rho}+\rho\dot{r}\geq 0$, using $r>0$ we can reduce this inequality
and $1$ to the conditions $\dot{r}>0$ and $ \dot{\rho}>-\frac{\rho\dot{r}}{r}$. Then 
\[ 
r\dot{r} +\rho\dot{\rho} > \frac{1}{r}(r^2-\rho^2)\dot{r}>0
\]
where we have used $3$ in the last inequality.
b) If $r\dot{\rho}+\rho\dot{r}< 0$, conditions $1$ and $2$ reduce to $\dot{\rho}<0$
and $\dot{r}>\frac{\rho-r}{r+\rho}\dot{\rho}$. Then
\[
r\dot{r} +\rho\dot{\rho}>\frac{1}{r+\rho}(-r^2+2r\rho+\rho^2) \dot{\rho}
\]
Now the bracket is negative precisely when $r>m_2\rho$. This together with 
$\dot{\rho}<0$ gives the result. $\;\Box$

\bibliographystyle{unsrt}
\bibliography{refs} 

\end{document}